\documentclass[twocolumn,twocolappendix]{aastex63}
\usepackage{graphicx}	
\usepackage{amsmath}	
\usepackage{amssymb}	
\usepackage{xspace}
\usepackage{natbib}

\newcommand{\hi}{\hbox{\ion{H}{1}}\xspace}

\newcommand{\numunit}[2]{\mbox{\ensuremath{#1\,#2}\xspace}}

\newcommand{\kms}{\ensuremath{\text{km}\,\text{s}^{-1}}\xspace}

\newcommand{\mk}{MeerKAT\xspace}

\newcommand{\rvirR}{\ensuremath{R_{200}}\xspace}

\newcommand{\mpc}{\text{Mpc}\xspace}
\newcommand{\zcl}{\ensuremath{z_{\rm cl}}\xspace}
\newcommand{\scl}{\ensuremath{\sigma_{\rm cl}}\xspace}
\newcommand{\mg}{\text{mag}\xspace}
\newcommand{\hwone}{Swarm\xspace}

\newcommand{\secref}[1]{Section~\ref{#1}}
\newcommand{\figref}[1]{Figure~\ref{#1}}
\newcommand{\tabref}[1]{Table~\ref{#1}}
\newcommand{\eqnref}[1]{Equation~\ref{#1}}

\graphicspath{{images/}}

\accepted{14 June 2021}
\submitjournal{AJ}

\shorttitle{Abell 2626 \& friends}
\shortauthors{Healy et al.}

\begin{document}

\title{Abell 2626 and friends: large and small scale structure}

\correspondingauthor{Julia Healy}
\email{healy@astro.rug.nl}

\author[0000-0003-1020-8684]{J. Healy}
\affiliation{Kapteyn Astronomical Institute, University of Groningen,
  Landleven 12, 9747 AV Groningen, the Netherlands} 
\affiliation{Department of Astronomy, University of Cape Town, Private Bag X3,
7701 Rondebosch, South Africa} 

\author[0000-0002-9895-5758]{S. P. Willner}
\affiliation{Center for Astrophysics \textbar\ Harvard \&
  Smithsonian,  60 Garden Street, Cambridge, MA 02138, USA} 

\author[0000-0001-9022-8081]{M. A. W. Verheijen}
\affiliation{Kapteyn Astronomical Institute, University of Groningen,
  Landleven 12, 9747 AV Groningen, the Netherlands} 

\author[0000-0002-5777-0036]{S.-L. Blyth}
\affiliation{Department of Astronomy, University of Cape Town, Private Bag X3,
7701 Rondebosch, South Africa} 




\begin{abstract}
  New MMT/Hectospec spectroscopy centered on the galaxy cluster A2626 and covering a ${\sim} 1.8\,\deg^2$ area out to $z \sim 0.46$ more than doubles the number of galaxy redshifts in this region. The spectra confirm four clusters previously identified photometrically. A2625, which was previously thought to be a close neighbor of A2626, is in fact much more distant. The new data show six substructures associated with A2626 and five more associated with A2637. There is also a highly collimated collection of galaxies and galaxy groups between A2626 and A2637 having at least three and probably
  four substructures. At larger scales, the A2626--A2637 complex is not connected to the Pegasus--Perseus filament.
\end{abstract}





\section{Introduction}
    \label{sec:intro}

        Large, wide-area spectroscopic surveys such as the Sloan Digital Sky Survey \citep[SDSS;][]{Strauss2002,Smee2012}, 2dF Galaxy Redshift Survey \citep[2dFGRS;][]{Colless2001}, and VIMOS Public Extragalactic Survey \citep[VIPERS;][]{Guzzo2013} have revealed the distribution of galaxies in the Universe. The emerging picture, now referred to as the ``Cosmic Web,''  is a complex and interconnected set of structures \citep{Lapparent1986, Bond1996}. These include megaparsec-long one-dimensional structures (``filaments''), thin two-dimensional sheets (``walls''), and vast empty spaces (``voids'') surrounded by walls. Galaxy clusters are located where multiple filaments intersect \citep[e.g.,][]{Aragon2010, Cautun2014, Malavasi2017}. 

        The early catalogs of galaxy clusters \citep{Abell1958,Zwicky1961} were constructed by identifying apparent over-densities of galaxies on photographic plates. \citet{Lucey1983} and \citet{Struble1987} estimated that as many as 25\%  of such identified clusters may be chance superpositions. The only way to confirm cluster members or even the existence of clusters at all is through extensive spectroscopy or at least multi-band imaging. Identifying cluster members can become particularly problematic when foreground and background galaxy overdensities overlap spatially, e.g., when the line of sight is along a cosmic filament.
        Early spectroscopy of Abell clusters
        (for a summary, see \citeauthor{Struble1999} and references therein) was usually limited to the brightest galaxies in the field. \citeauthor{Struble1999} commented that within their collation, as many as $5\%$ of the cluster redshifts may be incorrect and warned about the reliability of cluster redshifts determined from 1--3 galaxies. They endorsed the recommendation from \citet{Postman1986} that reliable cluster redshift measurements need to be made from at least five galaxies. 

        In the hierarchical structure-formation scenario, large structures such as galaxy clusters are built through successive merging and accretion of galaxies and groups of galaxies. The imprint of this accretion is told through the kinematic substructure within the cluster \citep[e.g.,][]{Dressler1988, Hou2012}. Identifying and studying substructure is an important step to uncovering how galaxy clusters form. Analysis of substructures can help to understand the local environments in which galaxies are located. The interaction between the cluster proper and the smaller group or substructure environment affects the evolution of the constituent galaxies. Processes such as galaxy--galaxy interactions are more likely to occur in the more densely populated environments, whereas processes such as ram-pressure stripping are more common in cluster environments where the intracluster medium (ICM) is denser. Understanding these effects may shed light on the causes of observed phenomena such as the morphology--density relation \citep{Dressler1980} in which dense environments such as clusters show a majority of red, quiescent galaxies whereas low-density environments have a majority of bluer star-forming galaxies. 

        The galaxy cluster at the center of this work is Abell~2626 \citep[officially ACO~2626 but hereafter A2626;][]{Abell1958}. A2626 is thought to be undergoing a merger between the main cluster and an accreted sub-cluster \citep{Mohr1996,Mohr1997}. Using $\sim$150 new redshifts, \citet{Mohr1996} found evidence for a potential sub-cluster southwest of the A2626 core. Newer spectroscopy from the WIde-field Nearby Galaxy cluster Survey \cite[WINGS;][]{Fasano2005,Cava2009} improved the redshift coverage for A2626, but \citet{Ramella2007} found no evidence of substructure in the cluster. These conflicting results are not surprising: different substructure algorithms are sensitive to different types of clustering (spatial or spectral) as well as the number of galaxies in the input catalog. 

        If A2626 is really undergoing a merger, it can act as a natural laboratory for the effects of mergers on constituent galaxies. At a distance of \numunit{250}{\mpc} and in a part of the sky that is not well surveyed, this cluster has not been extensively studied. Most of the recent interest in this cluster has focused on the intriguing ``kite'' source, visible at radio wavelengths at the center of the cluster \citep{Ignesti2018}. The cluster has been observed in X-rays, initially by \citet{Rizza2000}, and \citet{Wong2008} combined detailed X-ray observations with observations of the radio continuum to study the interplay between the physical processes that emit in both wavelength regimes. They found X-ray enhancements to the northeast and southwest of the cluster center and a significant jump in the radial gas temperature profile, which they suggested may be due to a past or ongoing merger event. The cluster is also home to six candidate ``jellyfish'' galaxies \citep{Poggianti2015}, extreme examples of galaxies undergoing ram-pressure stripping. One of the six, JW~100, a confirmed jellyfish galaxy, has been the center of a multi-wavelength (X-ray--radio) study to understand the interplay between the stripping processes and the stellar and gas components of the galaxy \citep[e.g.,][]{Poggianti2019,Moretti2019}. These studies have shown how important multi-wavelength data are to understand how the environmentally driven processes affect the evolution of galaxies, but equally  important is a thorough census of the galaxies in the environment.

        This paper presents new and deep spectroscopy of A2626 and its surroundings. The redshifts identify new members of A2626 and the nearby A2637 \citep{Abell1958} and reveal more of the large-scale structure in which these two clusters are embedded. The redshifts also identify a number of new structures in this region out to $z\sim 0.46$. 

        The new spectroscopy presented in this paper complements new radio observations centered on A2626 with \mk \citep{Healy2020b}.  Because  radio-frequency interference at the \mk site is so low, the new observations provide previously unattainable  neutral hydrogen (\hi)  measurements of galaxies in the cluster and its surroundings. In addition, \mk's wide bandwidth enables \hi observations of galaxies out to $z\sim0.46$. While \hi has yet to be directly detected at $z\ga0.376$ \citep{Fernandez2016}, optical redshifts can be used for \hi stacking---a technique where the \hi line spectra of a sample of galaxies are aligned and coadded to create an average spectrum with a higher sensitivity than the individual spectra \citep[e.g.,][]{Chowdhury2020, Healy2019}.

        The paper is organized as follows: \secref{sec:spectroscopy} presents the new spectroscopy. \secref{sec:lss} uses the new redshifts along with published ones to update the picture of the large-scale structure centered on A2626 at $z \la 0.46$. \secref{sec:clusters} provides updated measurements of  cluster sizes, and \secref{sec:substructure} describes substructures in and around A2626 and A2637. This paper uses throughout a standard, flat $\Lambda$CDM cosmology with $H_0 = 70$\,{\text{km}\,\text{s}$^{-1}$\,\text{Mpc}$^{-1}$}, $\Omega_M = 0.3$, $\Omega_\Lambda = 0.7$, and $h =0.7$.  All magnitudes are given in the AB system.

\section{Spectroscopy}
    \label{sec:spectroscopy}

        \subsection{Literature redshifts}
        \label{sec:a2626_litredshifts}

        WINGS \citep{Fasano2005} is a multi-wavelength imaging and spectroscopic survey of 77 galaxy clusters at Galactic latitude $|b| \geq 20^\circ$ selected from three flux-limited X-ray surveys. The main target of this work, A2626, is one of the clusters in the WINGS sample. WINGS used existing literature redshifts for the clusters in their sample in combination with their own follow-up spectroscopy survey \citep[WINGS-SPE,][]{Cava2009} to determine cluster membership. The targets for the WINGS-SPE were selected from the WINGS imaging survey \citep{Varela2009}, which for A2626, only extended to \numunit{r = 0.7}{\rvirR} where\rvirR is the radius inside which the mean density of the cluster is 200 times the critical density of the Universe. For A2626, \numunit{\rvirR = 1.64}{\mpc} = {25\farcm27} at $z = 0.0557$, \citealt{Cava2009}.) We are interested in the substructure of the cluster as a whole and understanding how the average gas content varies across the cluster based on local environment. For this, we need to be able to characterize the environment of the cluster beyond  \rvirR. 

        We supplemented the WINGS-SPE spectroscopy with redshifts by performing searches through the SDSS spectroscopic survey and the SIMBAD Astronomical database within a radius of 1\fdg5 from A2626. The radius of 1\fdg5 was chosen so that we could identify the known galaxies that are located in the area covered by the \mk\ \hi\ cube \citep{Healy2020b}, which covers \numunit{2\times 2}{\deg^2} centered on A2626. Despite falling into the SDSS spectroscopic footprint, as well as being part of the WINGS-SPE, the spectroscopic coverage across the entire \numunit{2\times 2}{\deg^2} field was very sparse and incomplete, especially for  $R > 0.7\,\rvirR$ from the cluster center. We therefore targeted A2626 for follow-up spectroscopy using the multi-object fiber-fed spectrograph Hectospec on the MMT.

        \subsection{MMT Spectroscopy}
        \label{sec:a2626_mmtobs}

        \subsubsection{Target selection}
        \label{sec:targetselection}

\begin{figure*}
            \centering
            \includegraphics[width=\linewidth]{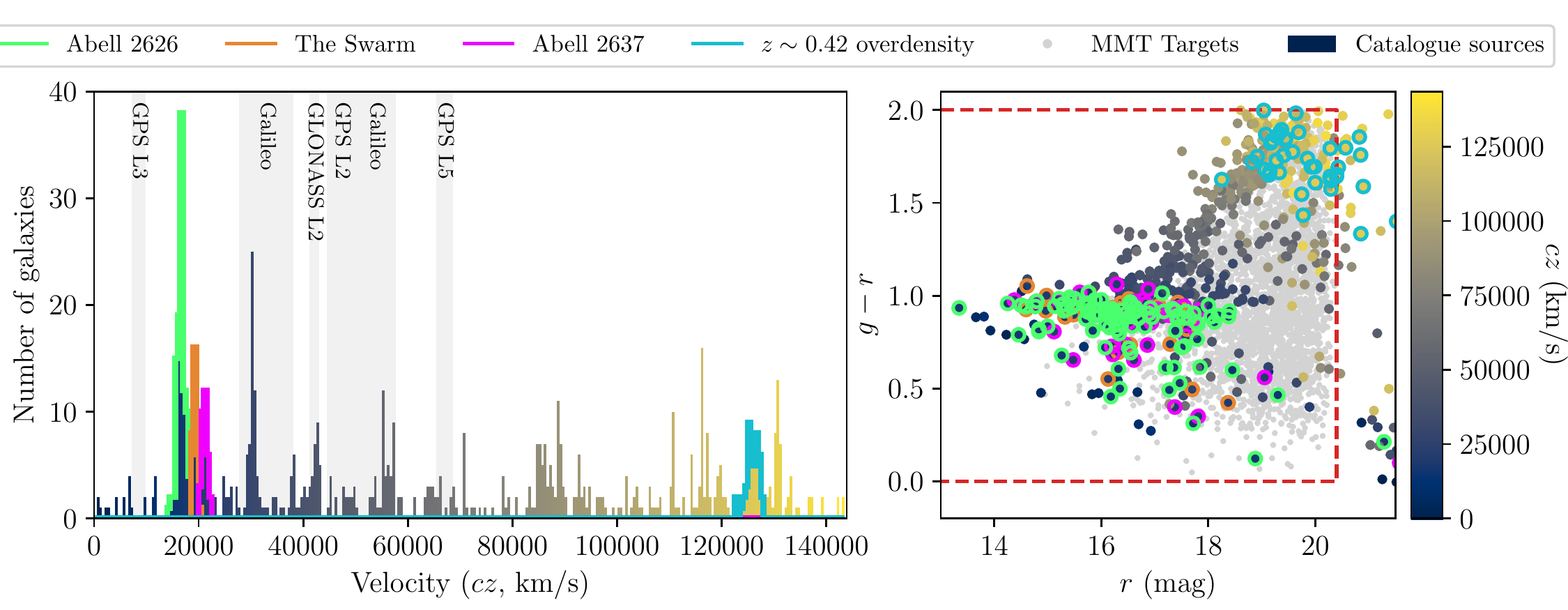}
            \caption{\textit{Left:} histogram showing the known redshifts of galaxies within a {1\degr} radius of A2626. The green, orange, and magenta outlines indicate the distribution of galaxies in A2626, a newly identified structure we are calling ``the \hwone'', and A2637 respectively. The blue outline indicates a background overdensity  at $z\sim 0.42$. Gray vertical bands mark the redshift ranges where radio-frequency interference makes it difficult to identify  galaxies in the \mk \hi data. Only overdensities outside these bands are colored. \textit{Right:} color--magnitude diagram for the galaxies in the field. The four overdensities are shown by open circles in the same colors as the histogram. The colors  of the filled circles represents the redshifts as indicated in the color bar to the right. The dashed red box represents the color and magnitude limits for the Hectospec target catalog.}
            \label{fig:mmt_target}
        \end{figure*}

        The galaxies targeted for Hectoscopec followup were selected from the SDSS photometric catalog. The first step was to identify all extended sources that did not already have a redshift. We classified  sources as extended if they satisfied the  criterion
        \begin{equation}\label{eqn:extendsources}
            r_{\rm psf} - r_{\rm petro} > 0.1\,\text{mag}\quad,
        \end{equation}
        where $r_{\rm psf}$ is the PSF magnitude, and $r_{\rm petro}$ is the Petrosian magnitude (an aperture-based magnitude that captures 80--100\% of the flux for most galaxies---\citealt{Blanton2000}). All sources that matched this criterion were cross-matched to the catalog of literature redshifts. Photometric sources within $5''$ of an existing redshift were assumed to be part of the same galaxy and removed from the target catalog. The distribution of literature redshifts (left panel in \figref{fig:mmt_target}) indicated three spectral overdensities at \numunit{cz \sim 20\,000}{\kms}, one of which corresponds to A2626. The highest-redshift overdensity was at \numunit{cz \sim 125\,000}{\kms} and centered around {0\fdg7} east-southeast of  A2626.  It is within the \mk primary beam  at that redshift. Based on the $g-r$ color, total $r$ magnitude \textbf{($r$ is the SDSS model magnitude unless otherwise specified)}, and locations of the noted four overdensities (see right panel of \figref{fig:mmt_target}), we further restricted the target catalog to:
        \begin{equation}
        \begin{array}{l}
             0 < g - r < 2,\\
            \numunit{r < 20.4}{\text{mag}}, \textrm{ and}\\
            {R < 0\fdg75}.
        \end{array}
        \end{equation}
        The color and magnitude limits are shown in \figref{fig:mmt_target}. We also imposed a brightness limit on the ``fiber magnitude,'' the magnitude within a $2''$ aperture, only including sources that had \numunit{r_{2''} < 21.5}{\text{mag}}. This  criterion enabled us to maximize the success rate of the observations. The color distribution of the prime target catalog of ${\sim} 2500$ sources is shown in \figref{fig:mmt_target}.

        \subsubsection{Observations}

        Hectospec \citep{Fabricant2005} is a fiber-fed spectrograph operating at the MMT in Arizona, USA.
        Its 300 fibers, each subtending a 1\farcs5 diameter on the sky, patrol a circle of radius 0\fdg5 with the constraint that the distance between fibers must be at least 20\arcsec. The 270~lpm grating gave a wavelength coverage 3725--9150~\AA\ with a spectral resolution 6~\AA.

        \begin{table}
        \begin{center}
        \caption{Hectospec Configurations}
        \label{t:configs}
        \begin{tabular}{ccccr}
        \hline\hline
        Config & R.A. & Dec. & Date & Exp.\ time\\
        &\multicolumn{2}{c}{J2000}& UT & minutes\\
        \hline
        c\_1 & 23:38:02.6 & 21:19:45 & 2019-09-03 & 105 \\
        c\_2 & 23:37:25.6 & 20:47:21 & 2019-09-04 & 105 \\
        b\_1 & 23:37:31.0 & 21:34:02 & 2019-10-24 & 45 \\
        b\_2 & 23:38:16.9 & 21:05:37 & 2019-10-24 & 54 \\
        a\_2 & 23:36:03.6 & 20:43:19 & 2020-06-20 & 45 \\
        d\_1 & 23:35:09.4 & 20:53:36 & 2020-06-22 & 114 \\
        e\_2 & 23:35:21.3 & 21:24:28 & 2020-09-11 & 60 \\
        f\_1 & 23:35:00.3 & 21:20:06 & 2020-10-07 & 120 \\
        e\_1 & 23:37:55.3 & 21:15:48 & 2020-10-08 & 45 \\
        \hline
        \end{tabular}
        \end{center}
        \end{table}

        The A2626 field was observed with nine Hectospec
        configurations designed to maximize the number of galaxies, and especially the number of bright galaxies, observed.  The plan was to observe five configurations with 45~minutes exposure time and four with 105~minutes, but in some cases longer exposures were obtained.
        Objects with SDSS $g$ fiber magnitudes $>$21 were preferentially assigned to configurations with longer exposure times.  This led to $\sim$280 galaxies being targeted twice.
        Each configuration was observed with three individual exposures to allow cosmic-ray removal.  Table~\ref{t:configs} shows positions, dates, and total exposure times of the configurations.
        (Configuration labels are arbitrary.)
        Each configuration measured about 250 targets and included 20--30 sky fibers.  Remaining fibers were unassigned because of position conflicts.  For the last three configurations observed, fainter targets were added in order to utilize more fibers.  These galaxies were observed only when no brighter source could be targeted.

        \begin{table}
        \caption{Template Uncertainties}
        \label{t:templates}
        \begin{tabular}{crcc}
        \hline\hline
        Template& $N$ & $\Delta cz$\tablenotemark{a}& $\Delta z$\\
        \hline
        \sc sptemp&1011&17&0.000057 \\
        \sc m31\_k\_temp&364&19&0.000063 \\
        \sc eltemp&294&25&0.000083 \\
        \sc hemtemp0.0&282&14&0.000047 \\
        \sc habtemp90&190&24&0.000080 \\
        \sc eatemp&69&29&0.000097 \\
        \sc m31\_f\_temp&46&23&0.000077 \\
        \sc m31\_a\_temp&11&31&0.000103 \\
        \hline
        \end{tabular}
        \tablenotetext{a}{Here and throughout this paper $c\equiv299792.5$.}
        \end{table}

        \begin{table*}
        \caption{MMT Redshifts}
        \label{t:mmtdata}
        \tiny
        \centering
        \setlength{\tabcolsep}{0.45em}
        \begin{tabular}{lccccccccccrrccl}
        \hline\hline
        \multicolumn{1}{c}{Name}
        &RA&Dec&petroMag\_g&PetroMag\_r&fiber2Mag\_g&cModel\_$g{-}r$&$Q$&$z$&$\Delta z$&alt&template
        &\multicolumn{1}{c}{$R$}
        &config&fiber&Comment\\
        \hline
        MMT\_1192  & 23:33:03.49& +21:11:03.6& 20.15& 19.26& 21.50& 0.91& 4&  0.101908& 0.000136&  &m31\_f\_temp&  6.6& f\_1& 096& \\
        MMT\_0441  & 23:33:03.74& +21:19:11.2& 17.82& 16.90& 19.91& 0.93& 4&  0.057972& 0.000074&  &    eltemp& 16.7& f\_1& 109& \\
        MMT\_1977  & 23:33:04.98& +20:58:04.5& 20.02& 18.53& 21.52& 1.61& 4&  0.248916& 0.000089&  &    eltemp& 17.3& d\_1& 169& \\
        MMT\_1185  & 23:33:06.15& +21:07:32.3& 19.01& 17.92& 20.85& 0.99& 4&  0.101535& 0.000041&  &  hemtemp0& 13.9& f\_1& 017& \\
        MMT\_1979  & 23:33:06.39& +21:00:30.7& 18.96& 18.04& 21.07& 0.92& 4&  0.099553& 0.000125& a&eltemp    &  9.2& d\_1& 161& \\
        MMT\_0443  & 23:33:10.49& +21:15:37.7& 19.14& 17.96& 20.58& 1.22& 4&  0.162603& 0.000058&  &m31\_k\_temp& 16.6& f\_1& 093& \\
        MMT\_0448  & 23:33:10.76& +21:17:27.2& 20.59& 19.14& 22.04& 1.46& 4&  0.224653& 0.000096&  &    sptemp& 13.3& f\_1& 106& \\
        MMT\_2430  & 23:33:11.54& +20:43:39.4& 20.42& 20.01& 21.67& 0.40& 4&  0.141438& 0.000018&  &  hemtemp0& 24.9& d\_1& 147& \\
        MMT\_2156  & 23:33:11.59& +21:25:35.4& 19.79& 18.94& 21.48& 0.86& 4&  0.192463& 0.000065&  &    sptemp& 17.5& f\_1& 116& \\
        MMT\_1196  & 23:33:11.86& +21:09:40.8& 20.35& 19.43& 21.69& 0.83& 4&  0.169041& 0.000087&  &    sptemp& 12.5& f\_1& 013& \\
        MMT\_0457  & 23:33:12.29& +21:23:58.7& 19.47& 18.44& 21.61& 0.93& 4&  0.265598& 0.000090&  &    sptemp& 13.5& f\_1& 112& \\
        MMT\_1198  & 23:33:13.62& +21:12:07.5& 21.49& 19.66& 22.81& 1.59& 4&  0.282104& 0.000147&  &    eltemp&  9.2& f\_1& 098& \\
        MMT\_1983  & 23:33:13.68& +20:55:35.8& 20.13& 19.46& 22.13& 0.62& 4&  0.142830& 0.000040& a&  hemtemp0&  9.0& d\_1& 164& \\
        MMT\_1989  & 23:33:14.20& +20:54:01.2& 19.22& 18.54& 21.74& 0.67& 4&  0.161680& 0.000061& a&  hemtemp0&  5.7& d\_1& 151& \\
        MMT\_0458  & 23:33:14.69& +21:20:10.2& 21.25& 19.97& 23.00& 1.17& 3&  0.284297& 0.000171&  & habtemp90&  7.1& e\_2& 094& \\
        MMT\_1980  & 23:33:14.70& +20:53:05.3& 19.38& 18.39& 20.90& 1.01& 4&  0.099837& 0.000053&  &m31\_k\_temp& 17.7& d\_1& 155& \\
        MMT\_2323  & 23:33:15.32& +21:05:11.0& 19.28& 18.62& 21.35& 0.68& 4&  0.146221& 0.000062&  &    sptemp& 18.6& d\_1& 176& \\
        MMT\_2319  & 23:33:15.91& +21:04:43.1& 21.08& 19.15& 22.24& 1.74& 4&  0.412846& 0.000156&  &    sptemp&  5.3& f\_1& 087& \\
        MMT\_2463  & 23:33:16.04& +20:52:21.4& 18.60& 17.59& 20.28& 1.01& 4&  0.099018& 0.000050&  &m31\_k\_temp& 20.6& d\_1& 157& \\
        MMT\_2462  & 23:33:16.28& +20:49:40.1& 17.56& 16.90& 19.40& 0.69& 4&  0.054066& 0.000033& a&  hemtemp0& 11.3& d\_1& 152& \\
        MMT\_1348  & 23:33:16.47& +21:29:37.0& 17.59& 17.21& 19.23& 0.37& 4&  0.038449& 0.000025&  &  hemtemp0& 15.6& e\_2& 107& \\
        MMT\_1194  & 23:33:16.56& +21:07:34.0& 20.20& 19.46& 21.58& 0.60& 4&  0.100518& 0.000096&  &    sptemp& 10.5& f\_1& 081& \\
        MMT\_0462  & 23:33:17.47& +21:22:12.6& 19.84& 18.96& 21.36& 0.88& 4&  0.057187& 0.000083&  &m31\_k\_temp& 10.3& f\_1& 026& \\
        MMT\_2464  & 23:33:17.48& +20:50:08.8& 18.83& 17.76& 20.41& 1.06& 4&  0.133259& 0.000077&  &    eltemp& 17.5& d\_1& 154& \\
        MMT\_1199  & 23:33:17.58& +21:09:45.3& 19.78& 19.00& 21.70& 0.64& 4&  0.019790& 0.000081&  &    sptemp& 10.5& f\_1& 015& \\
        MMT\_0451  & 23:33:17.58& +21:23:29.8& 18.13& 17.19& 19.84& 0.94& 4&  0.057222& 0.000051&  &m31\_k\_temp& 18.0& e\_2& 095& \\
        MMT\_2322  & 23:33:17.79& +21:03:31.9& 19.11& 17.96& 20.44& 1.14& 4&  0.132217& 0.000081&  &    eltemp& 17.5& d\_1& 271& \\
        MMT\_1981  & 23:33:17.82& +20:55:55.9& 19.66& 18.95& 21.71& 0.66& 4&  0.248011& 0.000057&  &    sptemp& 20.8& d\_1& 166& \\
        MMT\_0461  & 23:33:18.24& +21:16:19.8& 20.42& 19.32& 21.70& 1.10& 4&  0.165881& 0.000093&  &    sptemp& 13.4& f\_1& 091& \\
        MMT\_0463  & 23:33:18.74& +21:22:39.6& 21.54& 19.84& 22.75& 1.72& 4&  0.290004& 0.000146&  &    eltemp&  9.1& f\_1& 028& \\
        \hline
        \end{tabular}
        \raggedright
        \tablecomments{Magnitudes are from SDSS and are in AB units. An `a' in the alt column indicates the pipeline template choice was overridden. Only a portion of this table is shown here to demonstrate its form and content. A machine-readable version of the full table is available.}
        \end{table*}

        \subsubsection{Redshift measurements}
        MMT spectra were  reduced with the IDL Hectospec
        pipeline {\sc hsred} 2.1, originally written by R.~Cool (\url{http://mmto.org/~rcool/data_tools.html}).  The pipeline, a re-implementation of the one described by \citet{Mink2007}, combined the separate spectra while rejecting cosmic rays, subtracted the sky, and corrected the wavelengths to barycentric values.  The pipeline then cross-correlated each target spectrum with eight spectral templates representing a variety of galaxy spectra.  The template with the highest correlation coefficient was chosen as the initial redshift guess, and the location and width of the correlation peak gave the redshift and its statistical 1$\sigma$ uncertainty.
        There is an additional systematic uncertainty because the chosen template may not best represent the true best fit to the galaxy spectrum.  Table~\ref{t:templates} lists the galaxy templates, the number of times the pipeline chose each one, and the additional uncertainty due to the template choice.
        These uncertainties should be added in quadrature to the individual redshift uncertainties in Table~\ref{t:mmtdata}.

        All spectra were examined by eye and the correlations re-run with {\sc xcsao}, the original cross-correlation routine.  Redshifts from {\sc hsred} and  {\sc xcsao} were consistent within their uncertainties.  For about 10\% of targets, a template other than the initial guess looked to be a better representation of the spectrum, and the initial guess was replaced with the {\sc hsred} redshift for the better template.
        In most cases, the replacement template was chosen for better agreement with emission line velocities than the initial one.  For $\sim$10 targets, none of the {\sc hsred} redshifts matched the spectrum, and an {\sc xcsao} fit was used.  Most of these were spectra with large noise spikes that confused the pipeline but could be deleted manually.

        The visual examination gave each fit a quality ranking $Q$ to code whether the template had correctly identified the spectral features.  $Q=4$ designates an unambiguous redshift based on multiple features of high $S/N$\null.
        $Q=3$ designates a reliable redshift, but the number of features and their $S/N$ is insufficient for $Q=4$\null.  $Q=2$ designates a probable redshift but one with a reasonable chance of features being misidentified.  $Q=1$ designates some indication of a redshift, often a single, weak emission line or multiple features that could be absorption or could be noise.
        $Q=0$ indicates no spectral information, usually very low $S/N$.  Only $Q\ge3$ redshifts are used in this work, though others are included in Table~\ref{t:mmtdata} for completeness.

        \subsubsection{Data quality and success rate}
            \label{sec:mmtsucessrate}

\begin{figure}
                \centering
                \includegraphics[width=\linewidth]{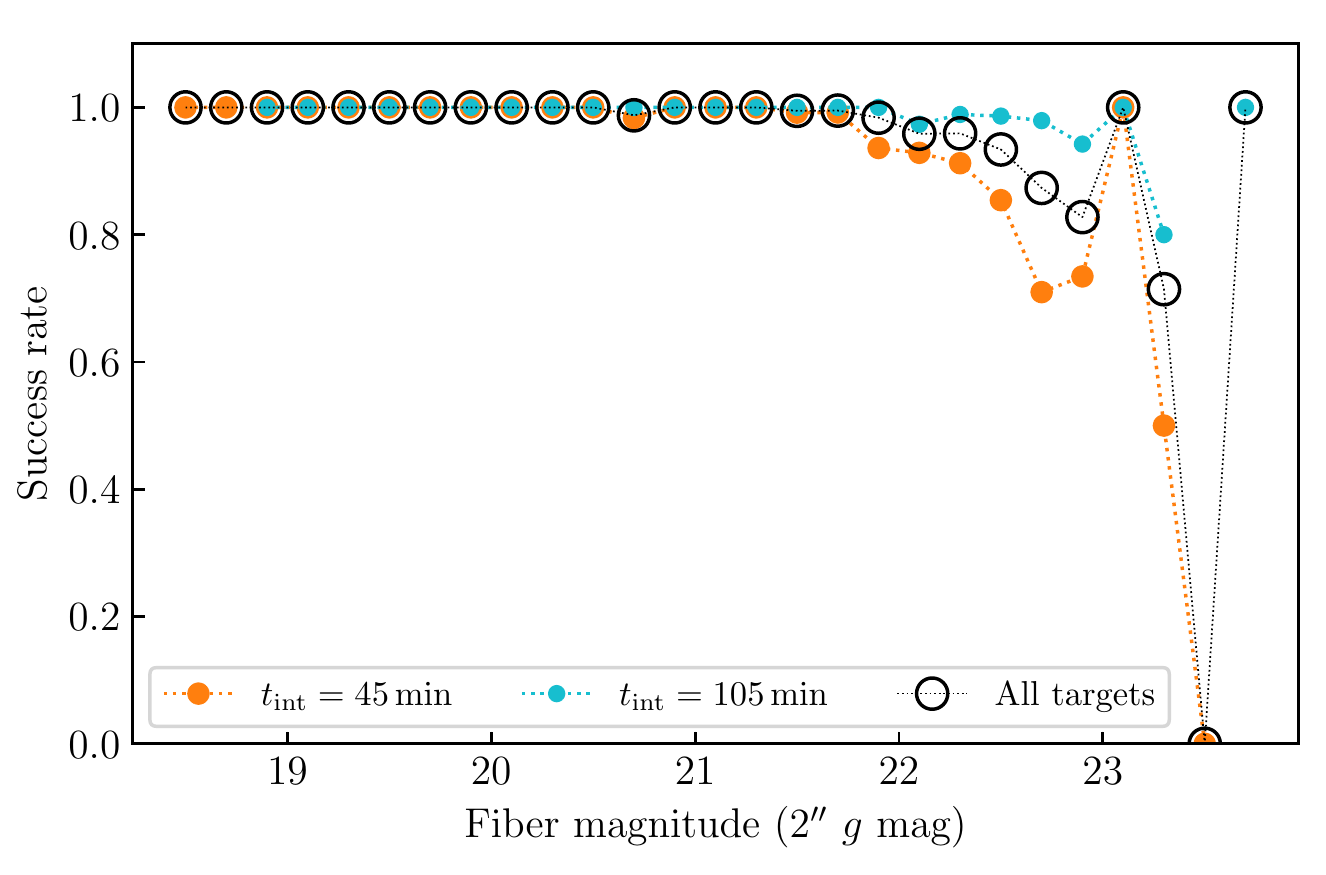}
                \caption{Fraction of targeted galaxies from which a redshift could be measured. The orange (blue) show the galaxies observed with \numunit{t_\mathrm{int} = 45}{\text{min}} (\numunit{t_\mathrm{int} = 105}{\text{min}}). The full target list is represented by the open black circles. The abscissa scale is SDSS $g$ magnitude in a 2\arcsec\ diameter (``fiber2Mag\_g").}
                \label{fig:mmt_success}
            \end{figure}

            The success rate is given as a function of the magnitude by
            \begin{equation}\label{eqn:success}
                S(m) = \frac{N_z}{N_\mathrm{targ}} (m)
                \quad,
            \end{equation}
            where $N_z$ is the number galaxies of magnitude $m$ for which a redshift could be measured, and $N_\mathrm{targ}$ is the number of galaxies that were targeted and a spectrum  obtained.
            \figref{fig:mmt_success} shows the success rate as a function of the fiber magnitude.

        \subsection{Spectroscopic Completeness}
            \label{sec:spec_completeness}
        \begin{figure}
        \centering
        \includegraphics[width=\linewidth,trim=0 18 0 0]
        {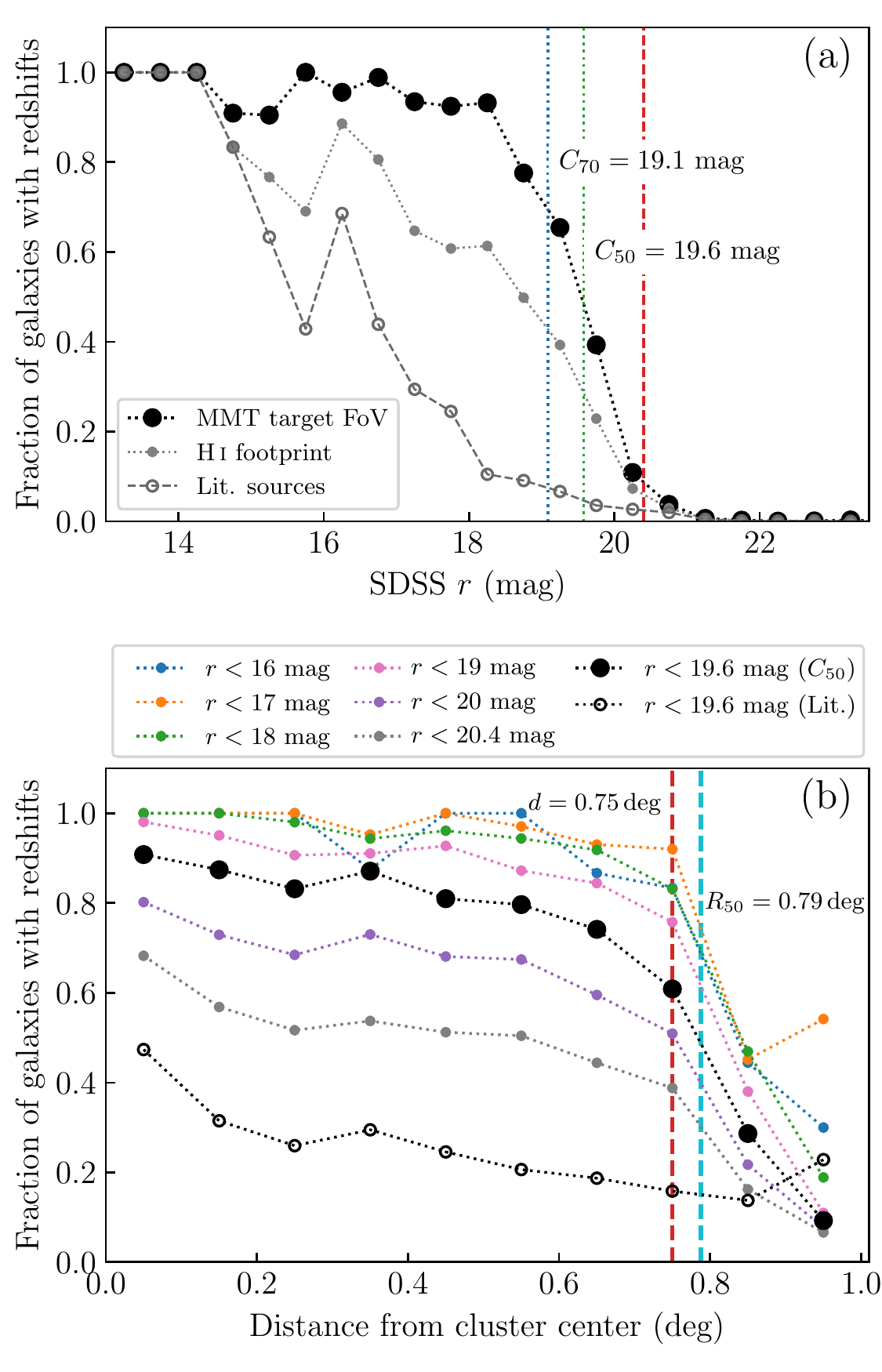}
        \caption{(a) Spectroscopic completeness as a function of SDSS $r$ across the A2626 field. Black filled circles show the completeness for the MMT target field with all available redshifts, gray filled circles show it for the entire \numunit{2\times 2}{\deg^2} \mk \hi footprint, and black open circles show the completeness for the MMT target field prior to this study. The blue and green dotted vertical lines indicate magnitudes for 70\% and 50\% completeness  for the MMT field. The red dashed line indicates the limiting magnitude for the MMT target catalog before the faint targets were added. (b) Spectroscopic completeness in annuli as a function of distance from the A2626 center. The black filled circles represent the completeness using all available redshifts for galaxies brighter than the $50\%$ completeness magnitude (\numunit{r < 19.6}{\mg}), and the black open circles represent the same prior to this study. Gray points represent all galaxies in the MMT target catalog (\numunit{r < 20.4}{\mg}). The blue, orange, green, pink, and purple points represent the completeness as a function of magnitude as indicated in the legend. The vertical dashed red line indicates the radius of the MMT survey region, and the vertical dashed cyan line indicates the radius at which the completeness for \numunit{r < 19.6}{\mg} galaxies falls below $50\%$.}
            \label{fig:a2626_completeness_oned}
        \end{figure}

        \begin{figure*}
            \centering
            \includegraphics[width=\linewidth]{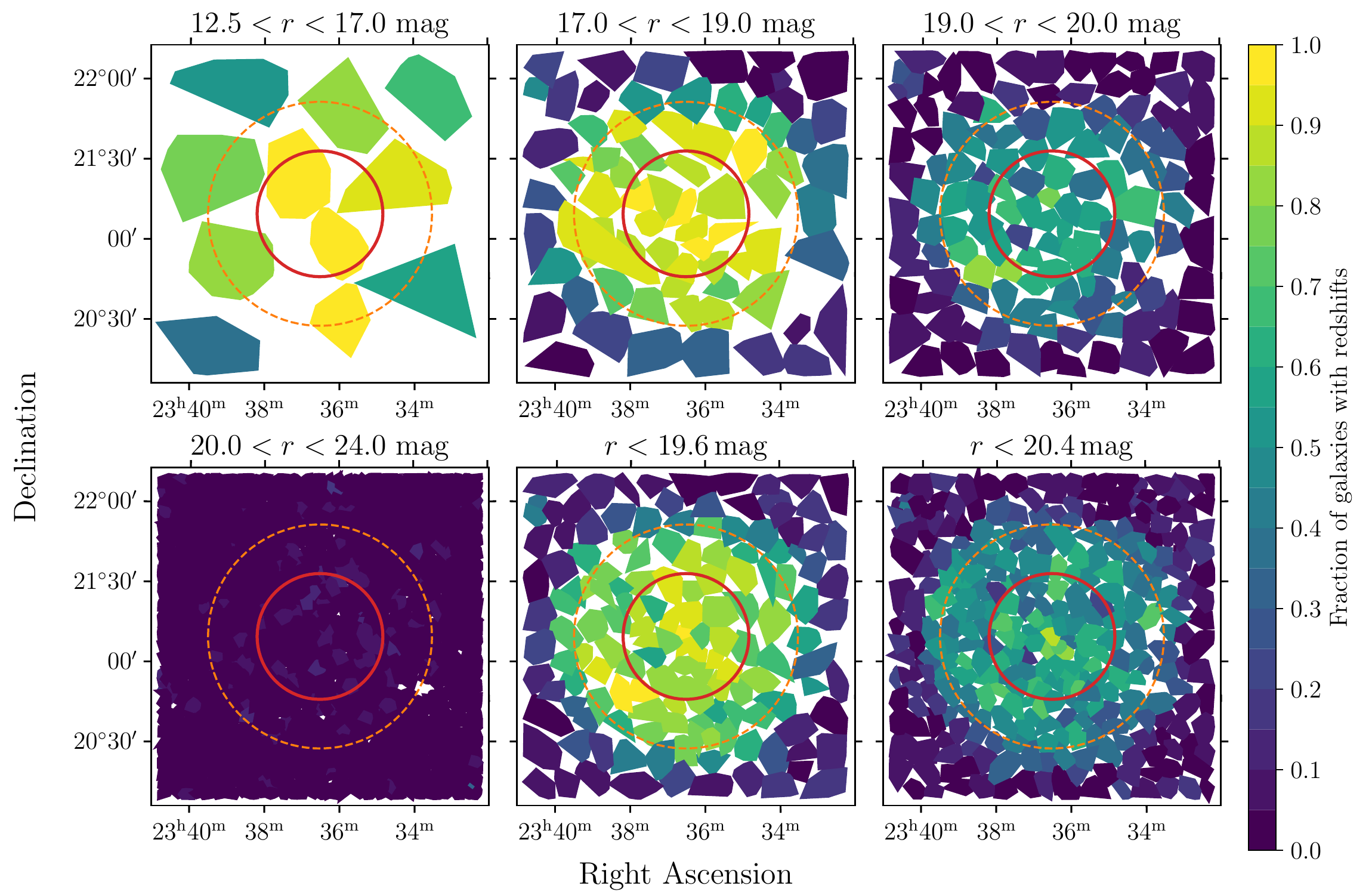}
            \caption{Spatial distribution of spectroscopic completeness across the A2626 field. The color of each tile indicates the completeness fraction $C(m)$ (\eqnref{eqn:completness}) of that subset of sources according to the color bar on the right. The solid red circle represents \rvirR of A2626 at the center of the field. The dashed orange circle indicates the outline of the MMT survey footprint. The irregular shapes of each tile are defined by the minimum area spanned by the positions of the galaxies represented by that tile.  In order to minimize the effects of stochasticity, each tile contains roughly the same number of sources.
            }
            \label{fig:a2626_completeness_cluster}
        \end{figure*}

        Understanding the completeness of our spectroscopic samples is important for understanding the reliability  of substructure and group identifications. Following the color and luminosity cuts outlined in \secref{sec:targetselection}, we examined the spectroscopic completeness across the entire \numunit{2\times 2}{\deg^2} region covered by the \mk \hi observations and centered on A2626. The completeness is defined in the same way as \citet{Cava2009}:
        \begin{equation}\label{eqn:completness}
            C(m) = \frac{N_z}{N_{\rm ph}} (m)
        \end{equation}
        where $N_{\rm ph}$ is the number of sources of magnitude $m$ in our target catalog, and $N_z$ is the number of sources of the same $m$ that have a redshift measurement (either from literature or MMT). \figref{fig:a2626_completeness_oned}a shows the spectroscopic completeness as a function of the SDSS $r$-band magnitude, which was used to select the targets for the MMT observations (\secref{sec:targetselection}). 
        \figref{fig:a2626_completeness_oned}b shows the completeness as a function of distance $d$ from the center of A2626.  The MMT spectroscopy improves the completeness fraction for sources fainter than \numunit{r > 15}{\mg} and at  \numunit{d < 0.8}{\deg}.

        While \figref{fig:a2626_completeness_oned}a provides the global overview at what magnitude the completeness begins to drops off, with a non-uniform spatial spectroscopic coverage it is important to understand how the completeness varies across the field. \figref{fig:a2626_completeness_oned}b shows the spectroscopic completeness determined in annuli with increasing radii from the center of the cluster.
        The spatial variation of the spectroscopic completeness in six magnitude bins
        is shown in \figref{fig:a2626_completeness_cluster}. The completeness is relatively uniform and above $50\%$ across most of the field.

\section{Large-scale structure}
    \label{sec:lss}

    \subsection{Large scale structure in the MMT volume}
    \label{sec:lssMMT}

        \begin{figure*}
        \centering
        \includegraphics[width=.95\linewidth]{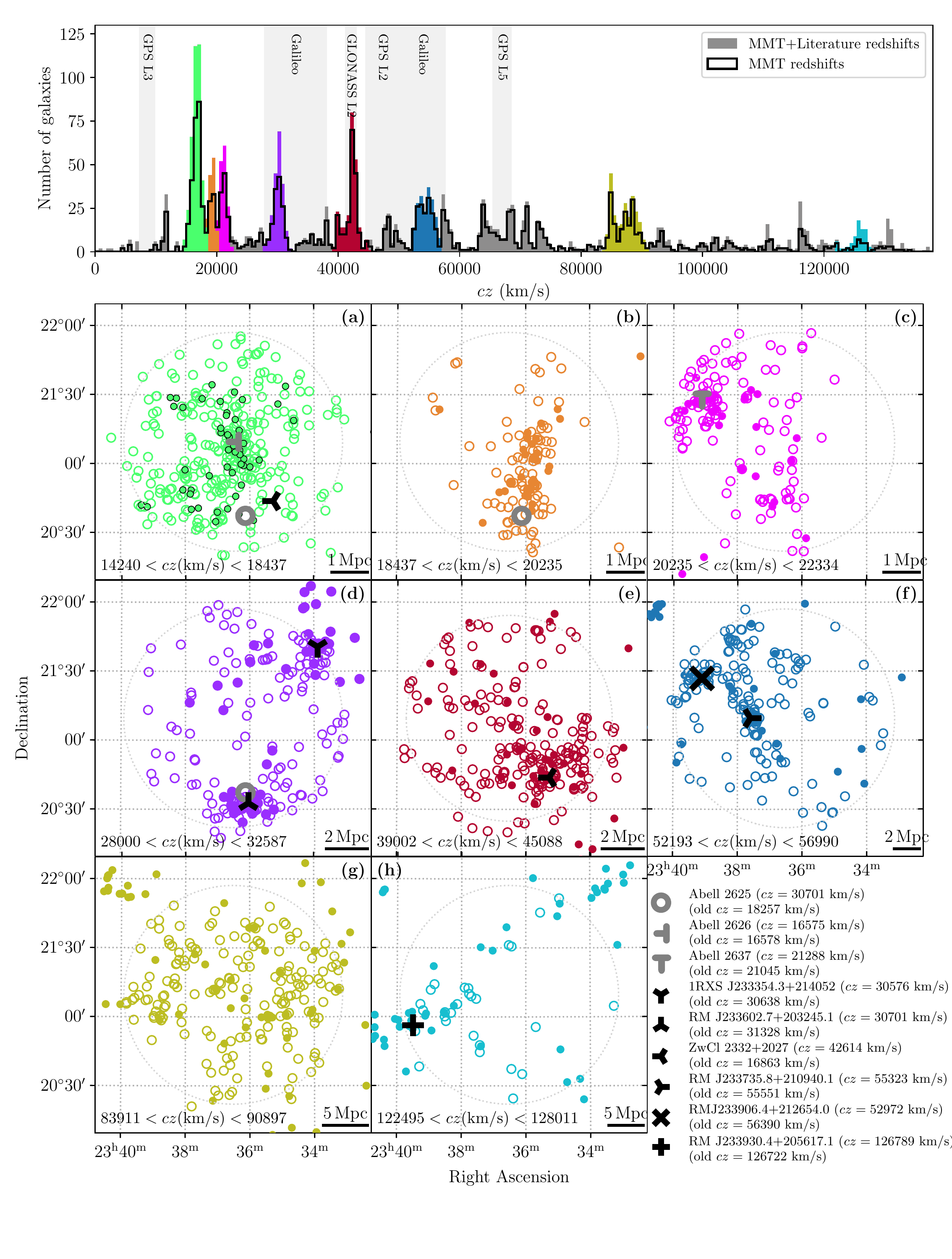}
        \vspace{-10ex}
        \caption{\textit{Top:} redshift histogram of all redshifts $z < 0.46$ (\numunit{cz < 137\,900}{\kms}) in a \numunit{2\times
            2}{\deg^2} region centered on A2626. The gray and colored histograms represent all known redshifts in this region, while the black open histogram represents our new MMT redshifts. \textit{Bottom:} sky distribution of eight of the spectral overdensities. Small circles represent galaxies with redshifts in the range indicated in each panel.  Open circles represent galaxies with MMT redshifts while filled circles represent galaxies with redshifts from the literature. Colors in each panel a--h are the same as the corresponding overdensity in the histogram above. The light gray dotted circles indicate the approximate outline of the union of MMT footprints  (\tabref{t:configs}). Known clusters in the region are indicated by markers as shown in the legend at bottom right.}
        \label{fig:MMT_hist_lss}
        \vspace{-1.5cm}
        \end{figure*}

        There are at least nine known clusters or overdensities within 0\fdg8 of A2626 and at \numunit{cz<138000}{\kms}.  These are listed in \tabref{tab:mkclusters}. 
        The three Abell clusters \citep{Abell1989} were found as surface overdensities of galaxies on the sky.  The five clusters identified as  ``RM'' were identified using the RedMaPPer algorithm \citep{Rykoff2014}, which searched for surface overdensities of red galaxies. ZwCL 2332+2027 was identified by \citet{Zwicky1961}, again as a galaxy surface overdensity, and  1RXS J233354.3+214052 was identified by \citet{Bohringer2000} from its X-ray emission. The MMT observations yielded 1858 new redshifts with \numunit{cz<138000}{\kms} and 20 with $cz > 138000$. With the addition of the MMT redshifts (\figref{fig:MMT_hist_lss}), many of these overdensities stand out as shown by  the histogram in \figref{fig:mmt_target}.

        The sky distributions of galaxies in eight velocity ranges are shown in \figref{fig:MMT_hist_lss}. The two clusters A2625 and ZwCl 2332+2027, which were identified from photographic plates, had previous cluster redshifts from only a few bright galaxies around the central position. Our new spectroscopy suggests that those galaxies are actually in the foreground, not members of either cluster. Updated cluster redshifts are shown in the figure legend and listed in \tabref{tab:mkclusters}.

        \subsection{Identifying A2625}
        \label{sec:a2625}
            
        \begin{figure}
        \centering
        \includegraphics[width=\linewidth]{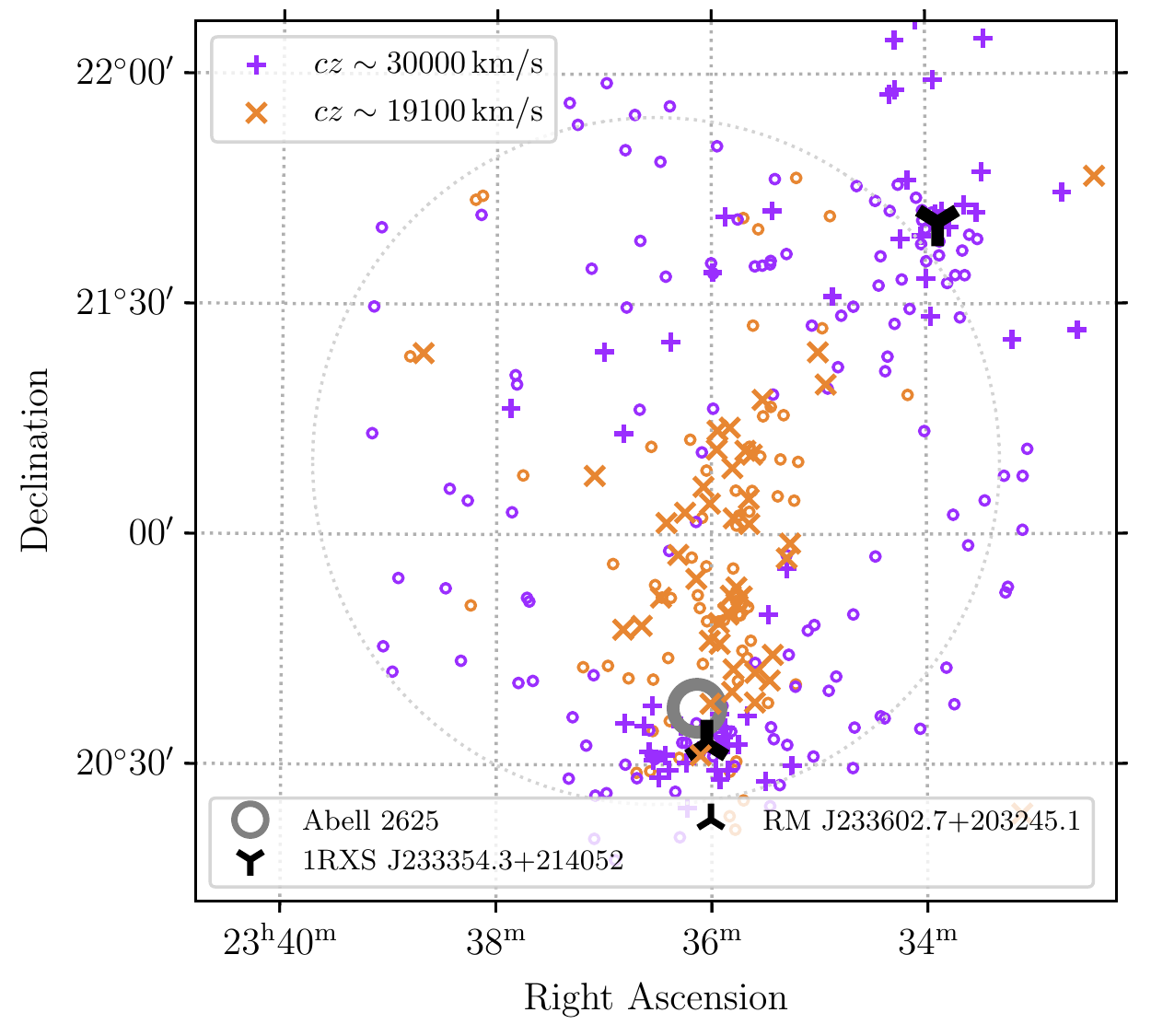}
        \caption{Sky distribution of the galaxies in the \numunit{cz \sim 19100}{\kms} overdensity (orange) and \numunit{cz \sim 30000}{\kms} overdensity (purple). The small open circles in the two colors represent the new MMT redshifts, while the orange `x' and purple `+' symbols represent measurements from literature. The black markers indicate the position of the two clusters associated with the two \numunit{cz \sim 30000}{\kms} overdensities. The position of A2625 is indicated by the open gray circle.}
        \label{fig:a2625_conundrum}
        \end{figure}
            
        The ACO catalog \citep{Abell1989} identified galaxy surface overdensities. These were later spectroscopically confirmed by measuring the redshifts of  the brightest galaxies in the field (sometimes as few as three galaxies, see \citealt{Struble1999}). A2625 was one of the clusters identified photographically, and the published cluster redshift was based the redshifts of three galaxies. \figref{fig:a2625_conundrum} shows a clear overdensity of galaxies at the A2625 location in both previously published and new redshifts.

        The previous redshift assigned to A2625 places it in velocity space between A2626 at \numunit{cz\sim16600}{\kms}
        and the overdensity around \numunit{19100}{\kms}.
        Based as it was on only three galaxies, that redshift was uncertain \citep{Struble1999}, and the three galaxies could be associated with the spectral overdensity at
        \numunit{cz \sim19100}{\kms} 
        (\figref{fig:MMT_hist_lss}b) rather than A2625. This redshift peak was identified as ``clump~B'' by \citet{Mohr1996}.  The many galaxies now identified in this peak show no significant spatial clustering but rather a more linear distribution stretching up the west side of A2626 (\figref{fig:a2625_conundrum}). There is also no X-ray emission associated with any of the localized clustering of galaxies in this overdensity.  Although the ACO position for A2625 is on the edge of the MMT survey area, there is enough coverage to have seen a cluster if it were there. Evidently the spectral overdensity around \numunit{19100}{\kms} is not associated with A2625, and we will henceforth refer to this overdensity as ``the \hwone.'' 

        \citet{Piffaretti2011} used {\it ROSAT} to identify X-ray emission of a galaxy cluster at A2625's position. There are two bright galaxies near the X-ray position and within an arcminute of each other: LEDA~97482 and LEDA~1630451. \citet{Owen1995} found \numunit{cz = 30130 \pm  90}{\kms} for LEDA~97482, and we found \numunit{cz = 17829 \pm 15}{\kms} for  LEDA~1630451  (\secref{sec:a2626_mmtobs}). The LEDA~1630451 redshift  places it in the \hwone. This galaxy's spiral morphology and detection in \hi \citep{Healy2020b} suggest that it is probably not close to the center of a cluster. LEDA~97482, on the other hand, is  a cD galaxy \citep{Yuan2016}. The X-ray emission and  cD galaxy are clear evidence of a cluster. Additional evidence  is the overdensity of red galaxies known as RM~J233602.7+203245.1  \citep{Bohringer2000}, now shown to be a  cluster in redshift space as well (\figref{fig:MMT_hist_lss}d).
        Taken together, 
        the clustering of galaxies, the X-ray emission at the center of the cluster, and the proximity to the original location of A2625 imply that A2625 and RM~J233602.7+203245.1 are the same cluster at \numunit{cz = 30519}{\kms}. \tabref{tab:mkclusters} gives cluster properties.

        \subsection{Large scale structure beyond the MMT volume}  
        \begin{figure}
        \centering
        \includegraphics[width=0.6\linewidth]{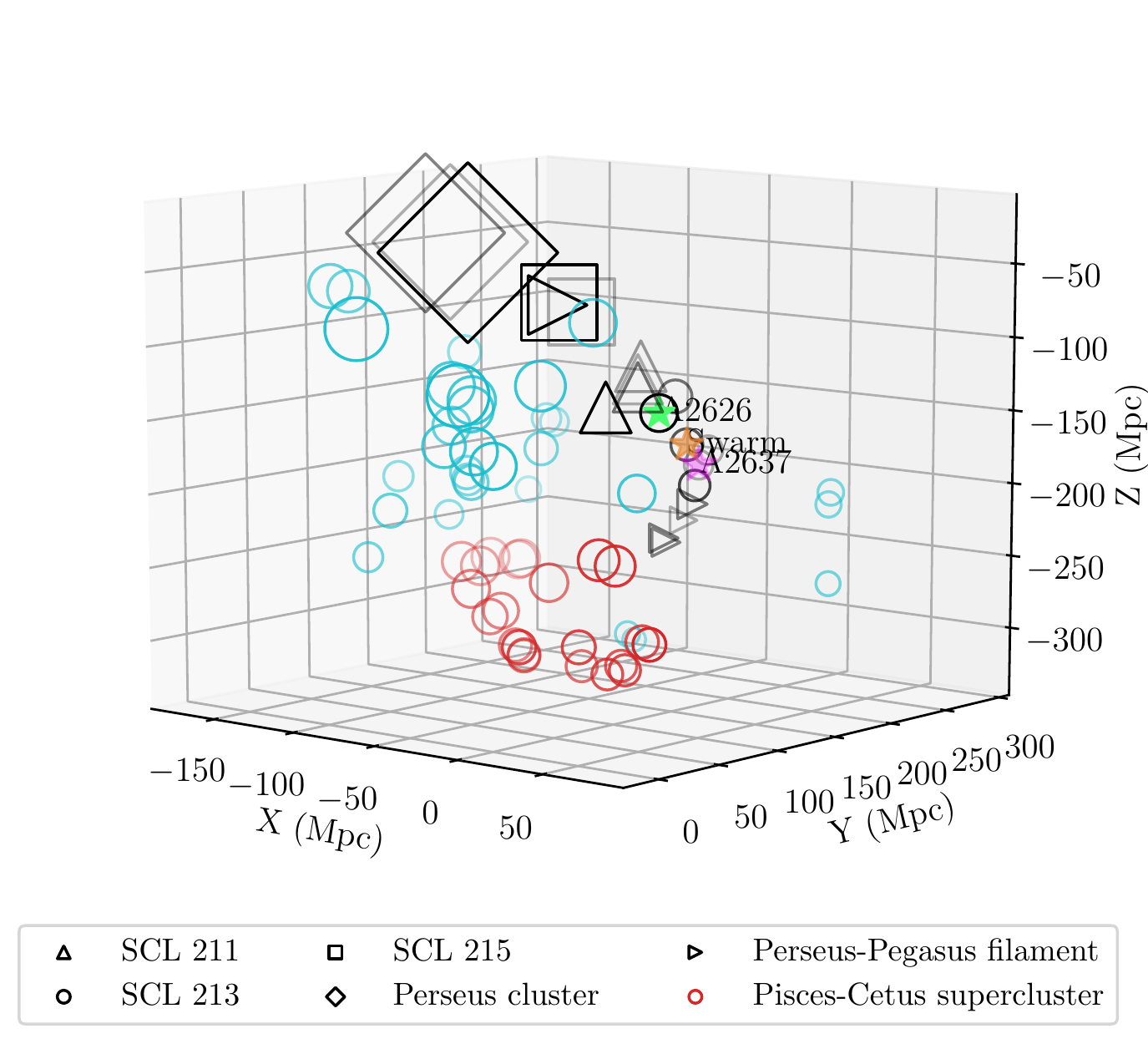}
        \includegraphics[width=0.6\linewidth]{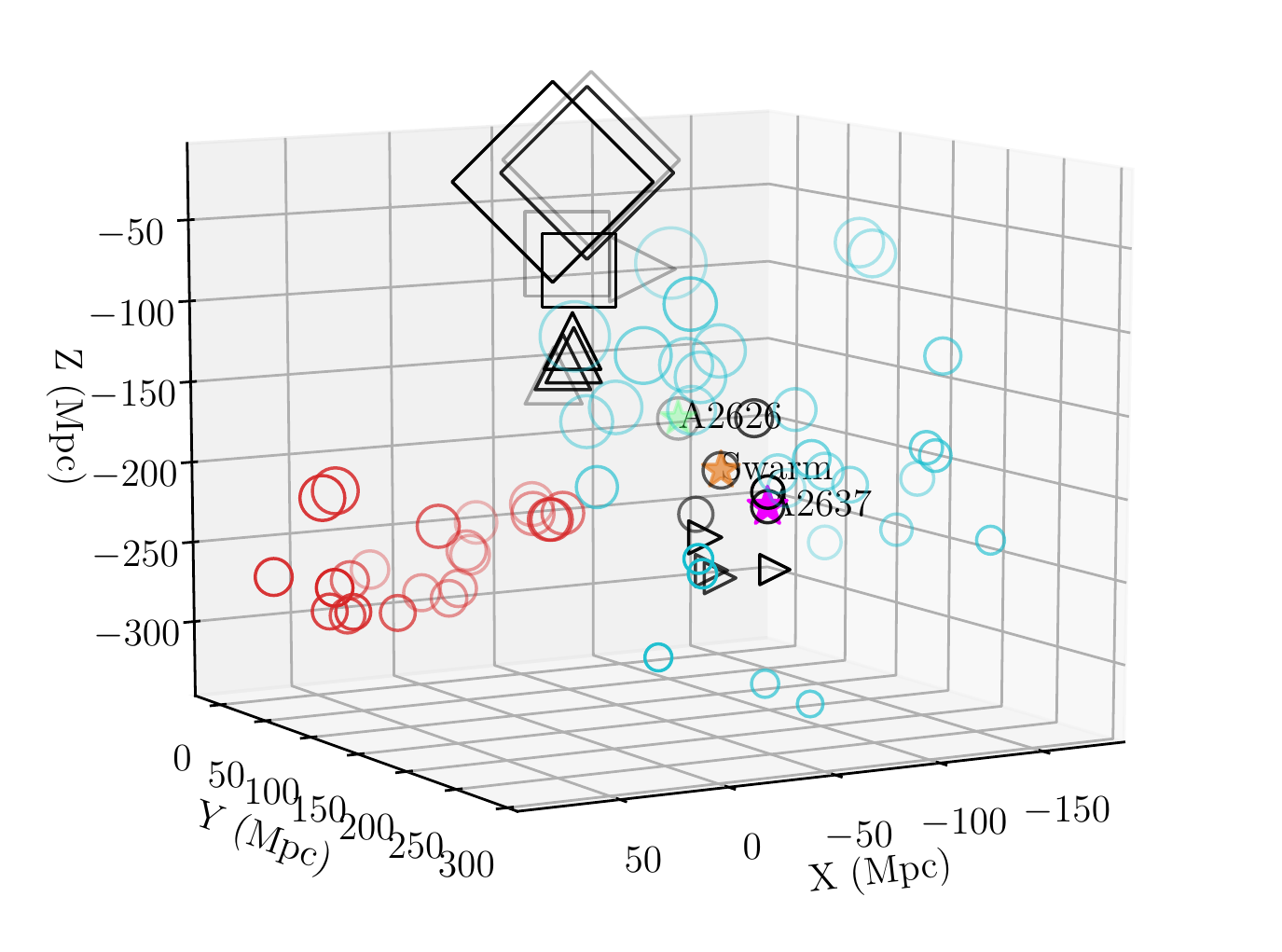}
        \caption{3-D distribution of the galaxy clusters in and around the Perseus--Pegasus filament. Coordinates are a Cartesian frame with the Galactic center at its origin. The top and bottom panels show two different azimuth angles. The filament is represented by the black shapes (diamonds = Perseus cluster; squares = SCL~211; upward triangles = SCL~215, circles = SCL~213, sideward triangles = other clusters in the filament). The red circles represent the Pisces--Cetus supercluster, and the blue circles represent other clusters in the field. The three clusters or overdensities in the low-redshift MMT volume are labeled and indicated by the green (A2626), orange (the \hwone), and pink (A2637) stars. \citet{Batuski1985} included the \hwone in their original version of this plot, labeling it A2625. The cluster marker sizes are scaled by the inverse of their redshifts and colored by structure: red is the Pisces--Cetus supercluster \citep{Tully1986,Tully1987}; black is the Perseus--Pegasus filament; and light blue is unassigned.  }
        \label{fig:pp_filament_3d}
        \end{figure}

        \begin{figure}
        \centering
        \includegraphics[width=\linewidth]{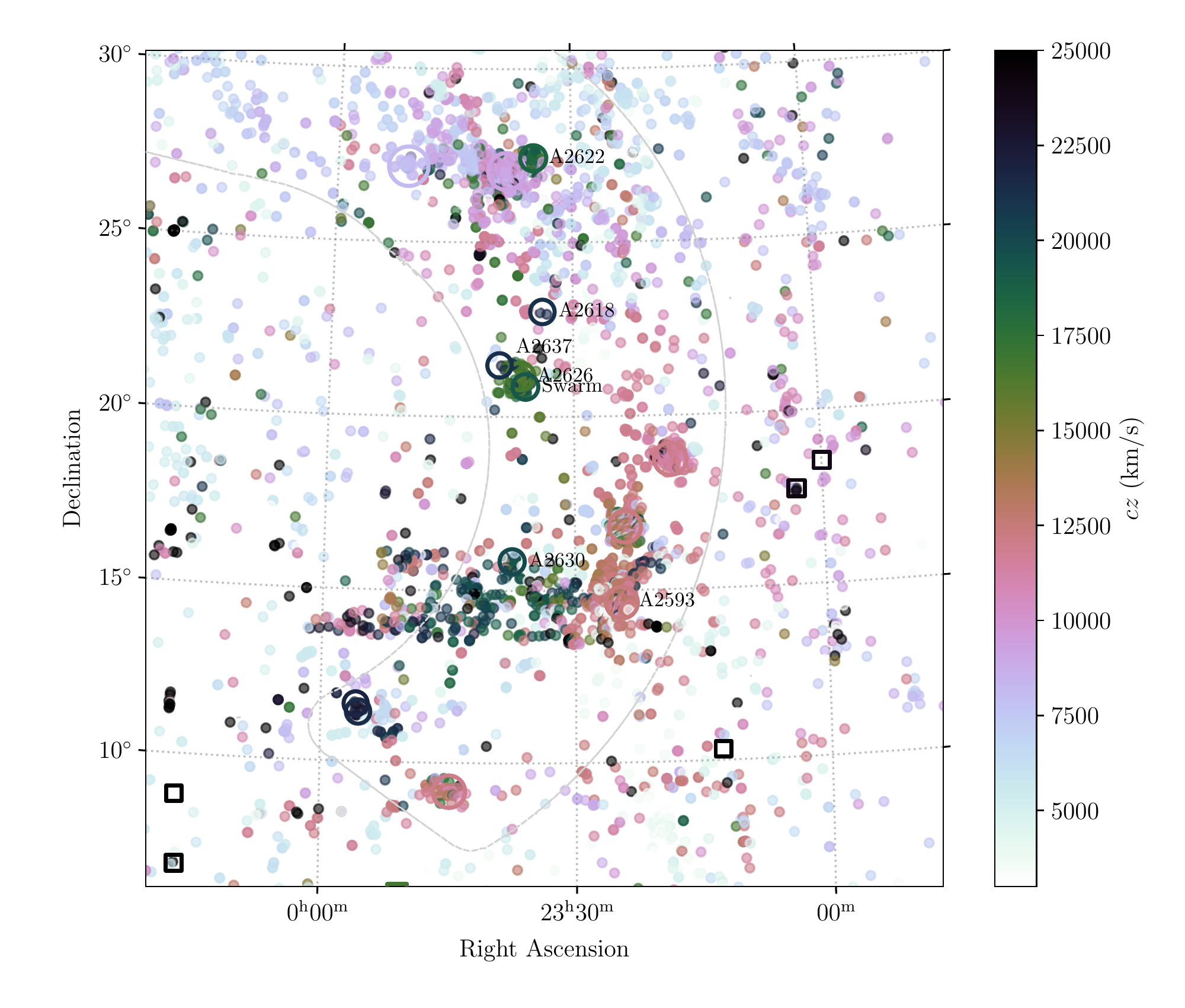}
        \caption{Sky distribution of the Perseus--Pegasus filament based on \citet[Fig.~3,][]{Batuski1985}. The filament is outlined by the light gray dashed line that resembles a hockey stick. The small colored points indicate galaxies in the region with redshifts from a combination of the CfA2, the Updated Zwicky catalogs, and targeted searches around the clusters in the {\sc simbad} database. The open circles represent Abell clusters identified by \citet{Batuski1985} as part of the filament, while the open squares represent clusters not in the filament. The size of the open markers is inversely proportional to the redshift of the cluster. The colors (as indicated by the color bar on the right) represent the recessional velocity of the object.}
        \label{fig:pp_filament}
        \end{figure}

        \citet{Batuski1985} identified a filament of galaxies extending over \numunit{425}{\mpc} from the Perseus--Pisces supercluster. \figref{fig:pp_filament_3d} shows a recreation of the \citet[][their Fig.~1]{Batuski1985} plot of the large scale structure using the most recent position and redshift information for the clusters in this region. \citet{Einasto2001} identified groups of galaxy clusters with \numunit{cz < 39000}{\kms} in a compilation of the ACO catalog and a sample of X-ray-detected clusters from \citet{Ebeling1998}. Using a friends-of-friends algorithm, \citeauthor{Einasto2001}\ identified the superclusters using linking lengths of 37--\numunit{54}{\mpc}. A number of these superclusters are located in the Perseus--Pegasus filament (\figref{fig:pp_filament}). One includes the Perseus cluster, which \citeauthor{Einasto2001}\ claimed is connected to SCL~211 and SCL~215 by ``free-floating'' clusters. Another \citeauthor{Einasto2001}\ supercluster is SCL~213, which is home to A2626, A2637, and four other clusters.

        Additional redshifts gathered since 2001 change the picture. From the  top panel of \figref{fig:pp_filament_3d}, the black symbols could represent a coherent filament. However, when one rotates the figure (bottom panel), there is a separation between SCL~211 (the triangles) and SCL~213 (the black circles). It appears that the closer part of the filament (containing Perseus, SCL~211, and SCL~215) is veering away from the direction of SCL~213.
        This separation becomes more evident from the sky distribution shown in \figref{fig:pp_filament}.  The filament is traced from the top left of the figure by the light blue through to the purple and then pink points coming down the center of the figure and is outlined by the gray ``hockey stick.'' On the outer edge of the hockey stick, we see a relatively smooth transition in colors, likely indicating that the galaxies and clusters are part of the same structure. Given the available data, there appears to be no connection between the outer and inner parts of the hockey stick, the inner part being dominated by dark green colors corresponding to significantly higher recession velocities. \citet{Batuski1985} described the filament as following the plane of the sky, twisting to a line-of-sight direction around A2593. That would predict a structure around \numunit{15000}{\kms} that would connect the plane of the sky portion to the line of sight. However, the transition of colors in \figref{fig:pp_filament} shows no such structure. Based on the available data, we conclude that the A2626 complex is separate from the Perseus--Pegasus filament.

        \begin{table*}
        \caption{Clusters and over-densities in the MMT footprint.}
        \label{tab:mkclusters}
        \centering{
        \begin{tabular}{lclcccr}\hline\hline
        Cluster    & RM\tablenotemark{g}             
        & \multicolumn{1}{c}{RA Dec} & $cz_{\rm cl}$  & \multicolumn{1}{c}{$\sigma_{\rm cl}$} & \multicolumn{1}{c}{$R_{200}$} & $N_z$     \\ 
        &  &   \multicolumn{1}{c}{(J2000)}  &    \multicolumn{1}{c}{\kms}      
        &  \multicolumn{1}{c}{Mpc}      &       \\ \hline
        A2626 & \nodata                     & 23:36:31.00 +21:09:36.3\tablenotemark{a} & 16576 & $660 \pm 26 $   & 1.59  &  163  \\
        The \hwone\tablenotemark{b}   &\nodata  & 23:35:55.26 +20:51:43.2                  & 19247 & $397 \pm 22 $   & 0.95  &  54   \\
        A2637 &\nodata                      & 23:38:57.80 +21:25:55.2\tablenotemark{c} & 21288 & $610 \pm 46 $   & 1.46  &  74   \\ 
        A2625 &  J233602.7+203245.1         & 23:36:08.20 +20:37:23.0\tablenotemark{d} & 30702 & $369 \pm 36 $   & 1.51  &  38   \\  
        1RXS J233354.3+214052& \nodata      & 23:33:53.00 +21:40:36.0\tablenotemark{e} & 30577 & $635 \pm 66 $   & 1.49  &  39   \\  
        ZwCl 2332+2027 &J233524.3+204336.1  & 23:35:18.00 +20:44:00.0\tablenotemark{f} & 42615 & $437 \pm 22 $   & 1.01  &  59   \\  
        \nodata & J233735.8+210940.1        & 23:37:35.80 +21:09:40.0\tablenotemark{g} & 55323 & $553 \pm 62 $   & 1.25  &  28   \\
        \nodata & J233906.4+212654.0        & 23:39:06.38 +21:26:54.0\tablenotemark{g} & 52972 & $458 \pm 35 $   & 1.04  &  30 \\
        \nodata  & J233930.4+205617.1           & 23:39:30.40 +20:56:17.0\tablenotemark{g} & \llap{1}26789&        \nodata       &    \nodata &   7    \\
        \hline
        \end{tabular}}\\
        {\raggedright
        \tablenotetext{a}{\citet{Cava2009}}
        \tablenotetext{b}{overdensity but not a cluster (\secref{sec:scl213}) originally identified as ``Clump~B'' by \citet{Mohr1996}. Position given is luminosity-weighted mean of galaxy positions.}
        \tablenotetext{c}{\citet{Patel2006}}
        \tablenotetext{d}{\citet{Piffaretti2011}}
        \tablenotetext{e}{\citet{Bohringer2000}} 
        \tablenotetext{f}{\citet{Zwicky1961}} 
        \tablenotetext{g}{\citet{Rykoff2014}}}
        \end{table*}
           
        \subsection{Large scale structure around A2626}
        \label{sec:scl213}

        \begin{figure*}
        \centering
        \includegraphics[width=\linewidth]{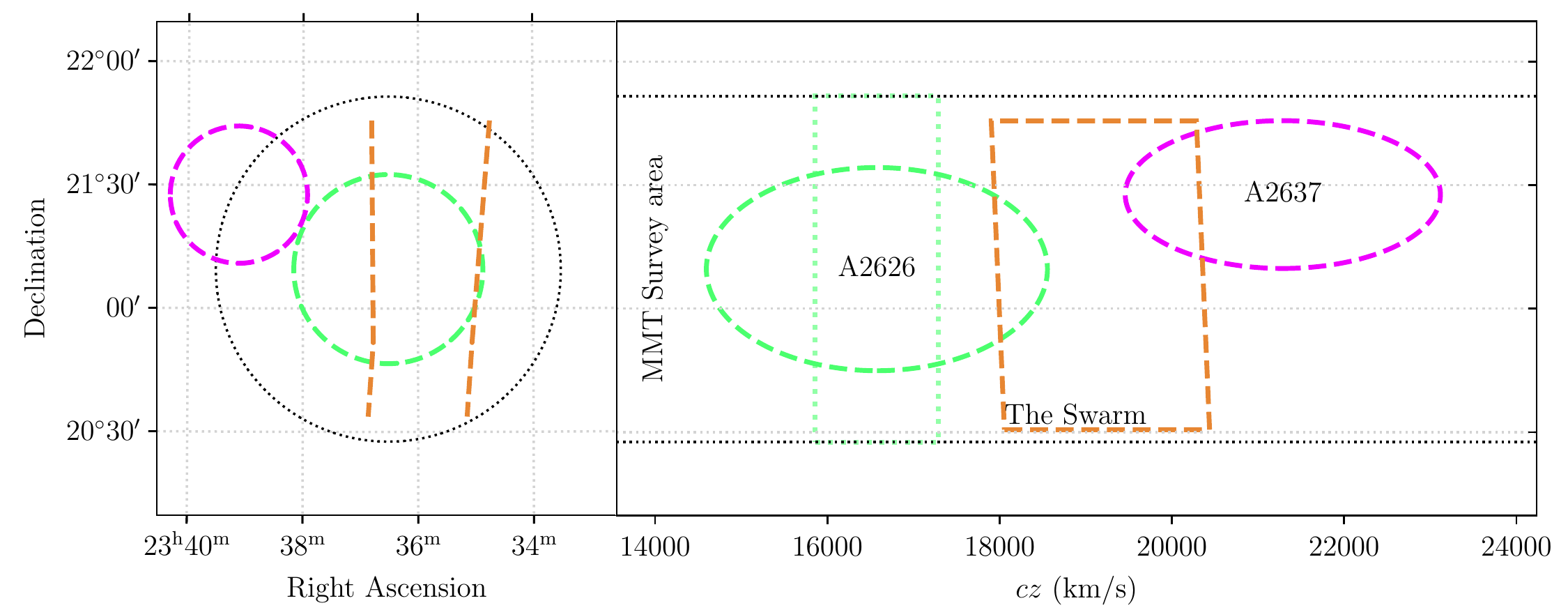}
        \caption{Schematic of how A2626, the \hwone, and A2637 fit together. The
          colored circles/ellipses represent A2626 (green) and A2637
          (pink). The orange rectangular shapes represent the \hwone, and the MMT
          survey area is indicated by the black circle and horizontal dashed
          lines. The wall in which A2626 is embedded is represented in the
          right panel by the light green vertical rectangle.}
        \label{fig:scl213}
        \end{figure*}

        Within a 2\degr\ radius around A2626 and at similar redshifts, there is one other Abell cluster: A2637 \citep{Abell1989}. These two clusters are evident in the redshift histogram for the field (\figref{fig:MMT_hist_lss}), where the two clusters stick out as distinct peaks with another overdensity, the \hwone, between them. Aside from the central region of A2626 \citep{Cava2009,Struble1999}, this region has not been extensively surveyed, and many of the existing measurements for the \hwone and A2637 have been based on a handful of redshifts and X-ray detections. 

        Despite \citet{Einasto2001} identifying A2626, the \hwone, and A2637 as part of SCL~213, the currently available redshift data for galaxies in this supercluster are sparse. Within our data, there are hints of how the three structures connect to the other clusters, but given the complexity of the cosmic web, it is not possible to show the links. Even with our limited field-of-view, there are still three interesting overdensities. 

        A2626 itself shows an extended distribution of galaxies (\figref{fig:MMT_hist_lss}a). If we exclude the virialized population within \numunit{1.5}{\mpc} of the cluster center, there is a smooth distribution of sources. This implies that the A2626 cluster is embedded within a wall as sketched in \figref{fig:scl213}. 

        A2637, another of the SCL~213 clusters in our field, is only half covered by the MMT survey footprint. Nevertheless, the data show a central, dense region of galaxies with a radial decrease in the density of sources. A2637 is well separated from A2626 in both position and velocity.

        The final member of SCL~213 in our field is what we refer to as the \hwone. \citet{Einasto2001} assigned this overdensity to SCL~213 under the name A2625. As discussed above, the \hwone is not A2625, but more important, it does not appear to be a cluster. There is no X-ray emission associated with any part of the \hwone. Moreover, the distribution of the the \hwone galaxies has no central, dense region. Instead the structure seems to be linear, almost like a filament. The \hwone structure starts around the same declination as A2626 but more to the east (\figref{fig:MMT_hist_lss}b) and stretches south to the limit of our field.  The highly linear distribution of the \hwone galaxies also stands out.

        \figref{fig:scl213} shows a schematic of the sky and redshift distributions of the three structures. The three overlap in the plane of the sky, but in redshift space, A2626 and A2637 are  distinct from each other. While the \hwone overlaps with A2626 in the plane of the sky, and both clusters in redshift space, we do not believe that the \hwone is connected to either cluster. \secref{sec:substructure_hw1} discusses this further.

        \subsection{Background clusters}
        \label{sec:bg}

        \tabref{tab:mkclusters} lists five clusters well separated from the A2626 complex.  One is 1RXS~J233354.3+214052 \citep{Rykoff2014} in the northwest  of the field (\figref{fig:MMT_hist_lss}d).  While it is at almost the same redshift as A2625, its projected separation is 8.0\,\mpc.
        ZwCl 2332+2027 (=RM J233524.3+204336.1) is shown in \figref{fig:MMT_hist_lss}e, and  RM J233735.8+210940.1 is shown in \figref{fig:MMT_hist_lss}f.  Updated parameters for both clusters are given in \tabref{tab:mkclusters}.  RM J233906.4+212654.0 is also visible in \figref{fig:MMT_hist_lss}f, and there is an additional galaxy concentration to its northwest.  These three clusters may be part of a larger structure with size scale $\sim$5\,\mpc.

        At still larger distances, there are three overdensities defined by clusters of red galaxies \citep{Rykoff2014}. These have too few redshifts to say much about them, but \tabref{tab:mkclusters} gives an updated redshift for RM J233930.4+205617.1, though based on only seven galaxies. The other two overdensities (J233502.7+210205.3 and J233905.9+210125.2) do not have enough redshifts to spectroscopically confirm the positions. \figref{fig:MMT_hist_lss}h shows a diffuse, linear structure, but the spectra are not deep enough to provide a good sample of galaxies  at this redshift.

\section{A2626 and friends}
    \label{sec:clusters}

        \begin{figure*}
        \centering
        \includegraphics[width=.99\linewidth]{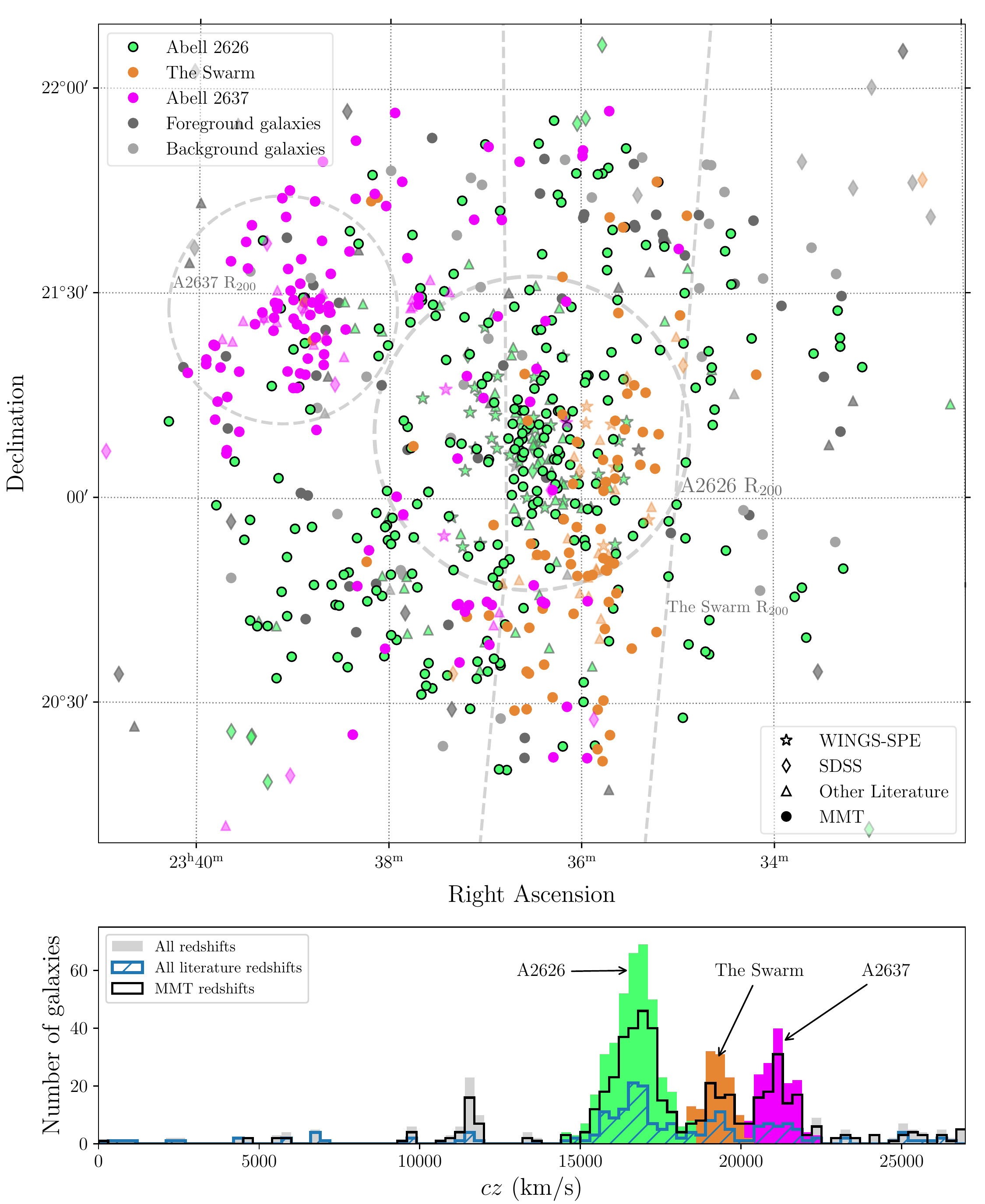}
        \caption{Top: sky distribution of all galaxies with $z < 0.09$ in a \numunit{2 \times 2}{\deg^2} region centered on A2626. The dashed gray circles indicates  \rvirR for the different overdensities as labeled. The different symbols represent the different sources from which the redshifts were obtained, and the colors represent the large scale structure to which the galaxies belong. Bottom: redshift histogram of all the galaxies in the upper panel. The solid histogram represents the entire catalog, while the open histograms indicated what portion of the total catalog comes from literature (hatched blue) and from the MMT observations (black).}
        \label{fig:a2626_skyplot_hist}
        \end{figure*}

        A goal of this work is to identify substructures within the clusters as well as  groups that may be a ``bridge'' between any of the clusters in the large-scale environment. This requires reliable measures of the cluster redshift (\zcl) and velocity dispersion (\scl). The redshift histogram in \figref{fig:a2626_skyplot_hist} gives estimates for \zcl and \scl for the three overdensities shown. In the rest frame of the overdensity, 
        \begin{equation} \label{eqn:restframev}
            v = c \dfrac{z - \zcl}{1 + \zcl}\quad,
        \end{equation}
        where $c$ is the speed of light, $z$ is the redshift of the galaxy, and \zcl is the redshift of the cluster. To determine \zcl, we selected all galaxies within a radius of \numunit{1.5}{\mpc} of the cluster center and with $-3\scl < v < 3 \scl$. We
        discarded all galaxies detected in \hi \citep{Healy2020b} because they do not typically trace the virialized population of a cluster \citep{Colless1996}. We applied the biweight location estimator to the redshifts of the sample of galaxies representing the virialized galaxy population \citep{Beers1990} to determine \zcl. Using the updated \zcl, we recalculated the rest-frame velocities, $v$, for each galaxy using \eqnref{eqn:restframev}. We determined \scl by fitting a Gaussian to the histogram of rest-frame velocities, fixing the mean of the Gaussian to the cluster redshift.
        Our \numunit{\scl = 660 \pm 26}{\kms} for A2626 is fully consistent with the \numunit{679 \pm 60}{\kms} measured by \citet{Cava2009}. 

        \begin{figure*}[t!]
        \centering
        \includegraphics[width=\linewidth]
        {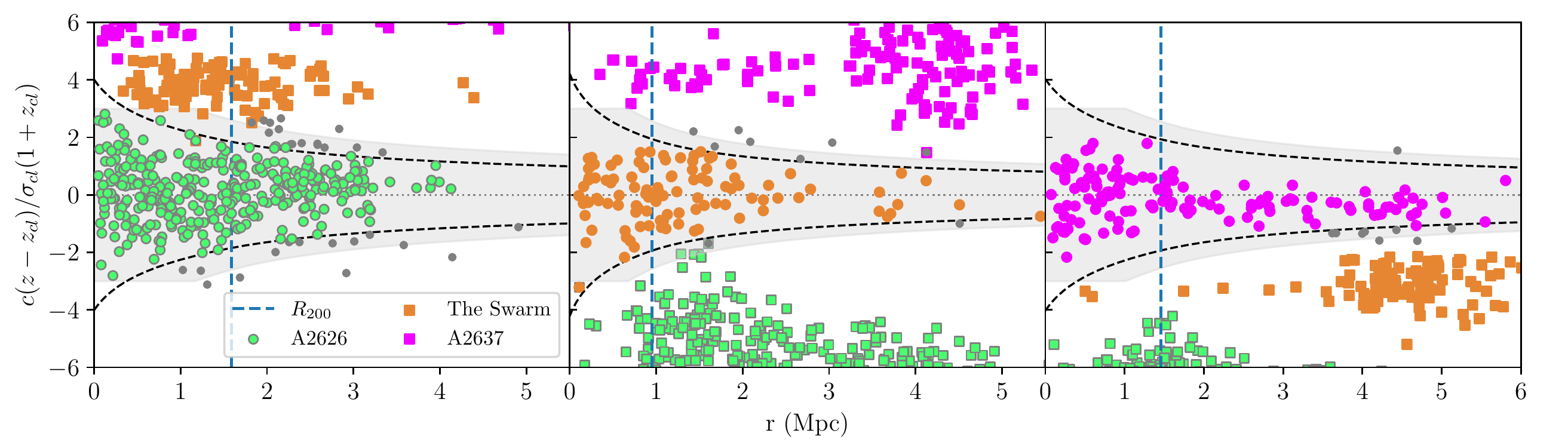}
        \caption{Phase space diagram for A2626 (left), the \hwone (middle), and A2637 (right). The black dashed lines indicate the escape velocity for each overdensity. This was calculated using the formalism from \citet{Jaffe2015}, assuming a concentration index $C=6$ and the mass enclosed by \rvirR from \eqnref{eqn:m200}. Galaxies located within the escape velocity trumpet for each overdensity are represented by the colored data points. The gray circles represent galaxies that satisfy $-3\scl < v < 3\scl$ (Eq.~\ref{eqn:restframev}) but are not located within the escape-velocity trumpet for the particular overdensity. The vertical dashed blue line indicates \rvirR for each overdensity. The gray shaded region shows the galaxies used in the DS test for each overdensity.}
        \label{fig:a2626_phasespace}
        \end{figure*}

        The cluster velocity dispersion determines \rvirR, the radius inside which the mean density of the cluster is 200 times the critical density of the universe. From \citet{Finn2005} for $h=0.7$,
        \begin{equation} \label{eqn:r200}
            \begin{array}{ll}
            \rvirR = & 2.47 \left(\dfrac{\scl}{\numunit{1000}{\kms}}\right)\times\\
            &\quad ({\Omega_\Lambda + \Omega_M (1 + \zcl)^3})^{-1/2} \,
            \mathrm{Mpc}. 
            \end{array}
        \end{equation}
        The calculated \zcl, \scl, and \rvirR for the three overdensities are in \tabref{tab:mkclusters}. the \hwone is not a
        cluster nor is it a virialized system, but the numbers are useful for determining which galaxies may belong to the overdensity. 

        The newly calculated \zcl, \scl, and \rvirR for A2626, A2637, and the \hwone show how galaxies in these systems relate to each other in angular distance and velocity. \figref{fig:a2626_phasespace} shows the phase space plots for the three systems.  The virial mass $\mathrm{M_{200}}$ was calculated from \rvirR via
        \begin{equation}\label{eqn:m200}
        {M}_{200} = \frac{4}{3} \pi \rvirR^3  200\rho_c \,.
        \end{equation}

        \figref{fig:a2626_phasespace} shows that the three overdensities are distinct systems. The large velocity separation coupled with the distance between the three systems' centers of mass suggests that they are not even interacting with each other. the \hwone panel of \figref{fig:a2626_phasespace} reinforces that the \hwone is not a cluster because galaxies do not fill the flare of the trumpet at  small radii. The A2637 trumpet also does not fill up at small radii, but this could be due to spectroscopic incompleteness as A2637 is on the northeastern edge of the MMT survey area.

\section{Substructure in A2626}
    \label{sec:substructure}

        \citet{Mohr1996} identified three subcomponents with distinct velocities within \numunit{2.1}{\mpc} of A2626. Their Groups~A, B, and~C match our definitions of A2626, the \hwone, and A2637, respectively. Applying the Dressler-Shectman \citep[DS,][]{Dressler1988} test to their 159 redshifts, \citet{Mohr1996} found no substructure within any of the three groups. This is perhaps not too surprising as the DS test (like many substructure finding methods) is sensitive to the number of redshifts used. More recently, using new data from the WINGS survey, \citet{Ramella2007} applied a non-parametric clustering algorithm to A2626 and also found no significant substructure. This is also unsurprising given that the WINGS spectroscopy is limited to $R < 0.7$\,\rvirR and redshifts of 76 galaxies. Early work on clusters such as the Coma cluster found no significant substructure within the cluster \citep{Dressler1988}, but later works with more redshifts found the Coma cluster to contain a significant amount of substructure \citep{Adami2005, Healy2020}.

        \begin{figure*}
        \centering
        \includegraphics[width=\linewidth]{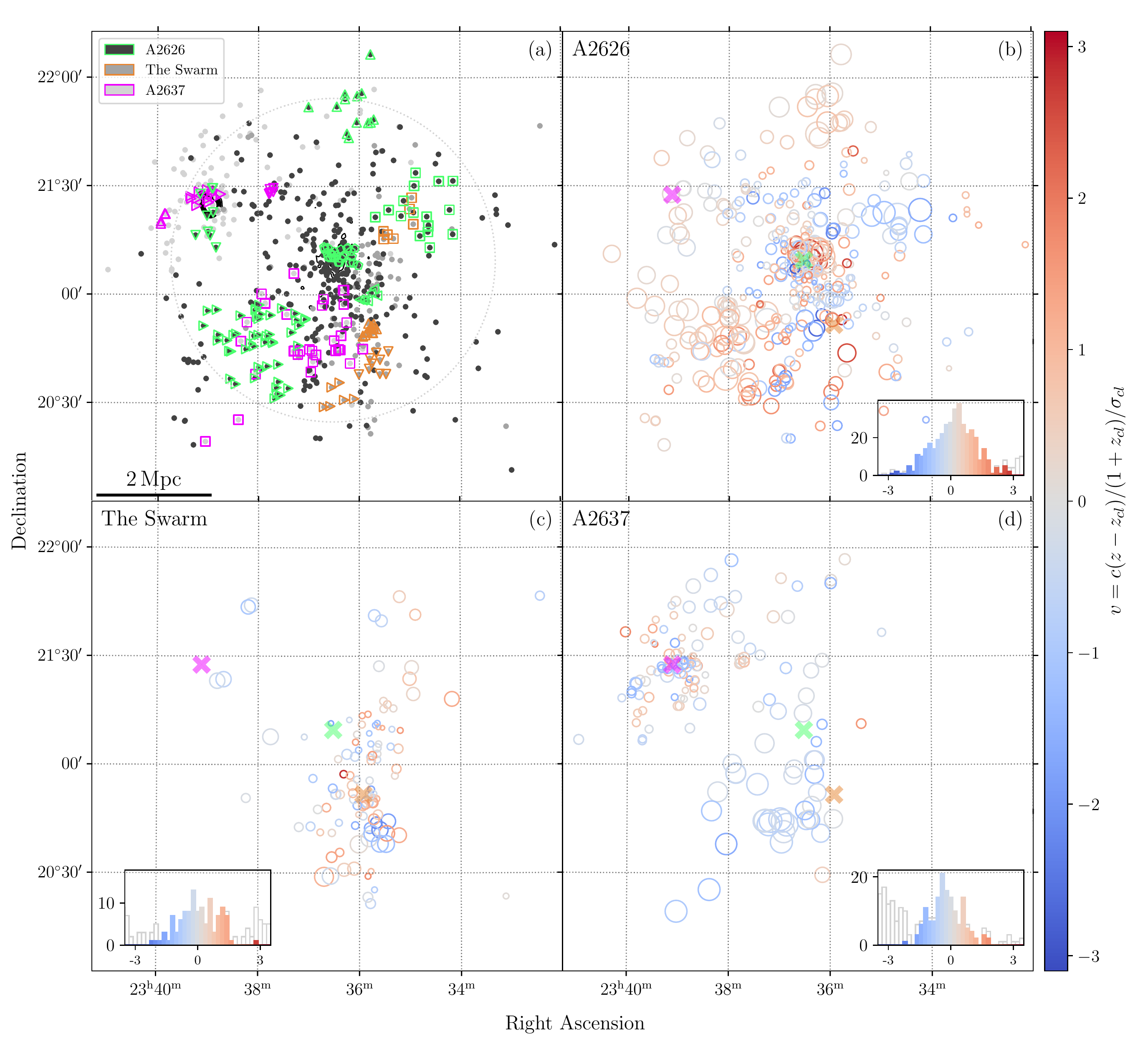}
        \caption{Sky distribution of galaxies in the overdensities. Colors in each panel mark related substructures. Panel (a) shows all three main structures A2626, the \hwone, and A2637 as marked by the colors of the open symbols. Shapes of the open symbols indicate the sub-structure group to which each galaxy belongs. The black, gray, and light gray points represent the galaxies in A2626, the \hwone, and A2637 respectively  not associated with any substructure. Panels (b), (c), and (d) show the  results from the DS test for A2626, the \hwone, and A2637 respectively. The size of each symbol is scaled by $e^{\delta_i}$ where $\delta_i$ is calculated by \eqnref{eqn:dstest} for $N_{\rm nn} = 10$. The large crosses mark the three main overdensity centers. The histograms in the lower corners of the panels show the velocity distribution of the sources in that panel with colors indicating velocity relative to the mean velocity of the overdensity.  The same colors show the velocity of each galaxy. Horizontal units are \scl as given in \tabref{tab:mkclusters}.} 
        \label{fig:a2626_dstest}
        \end{figure*}

        \subsection{Identifying substructures}
        \label{sec:a2626_identifysubstructure}

        \begin{figure*}
        \centering
        \includegraphics[width=.96\linewidth]{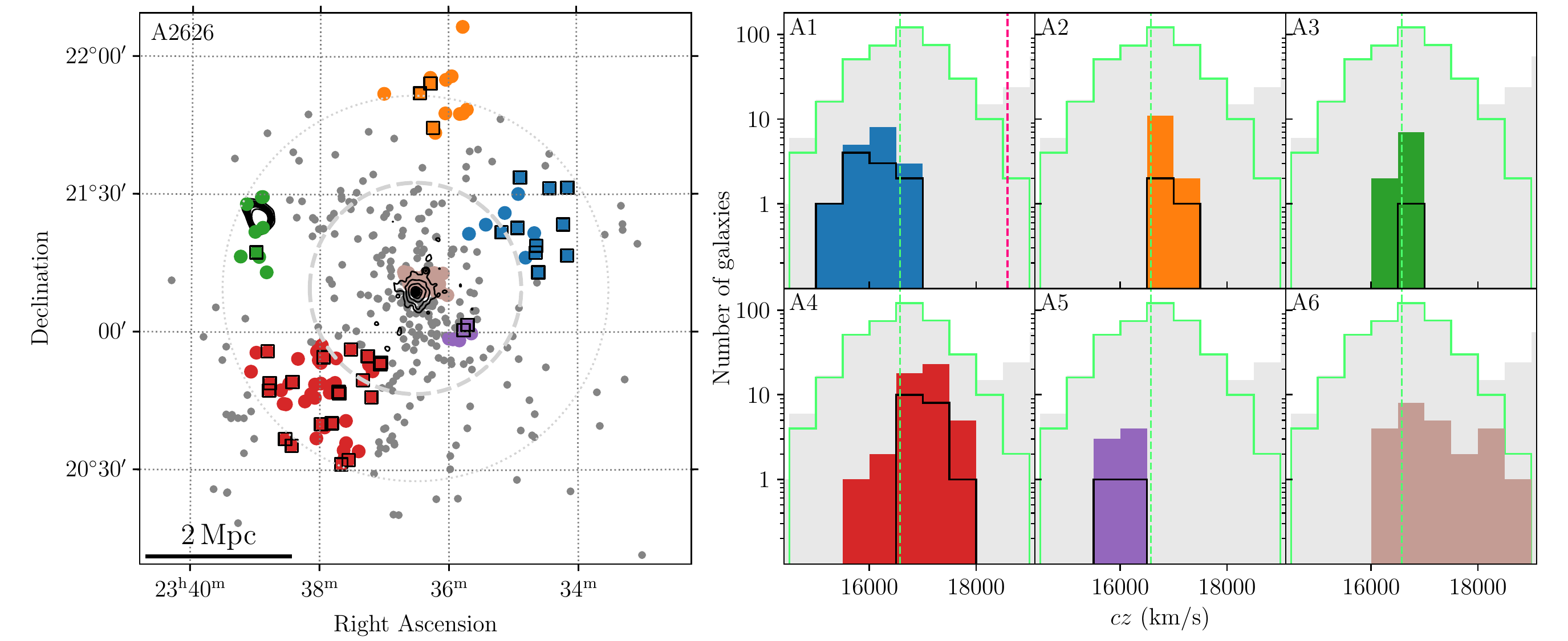}
        \includegraphics[width=.96\linewidth]{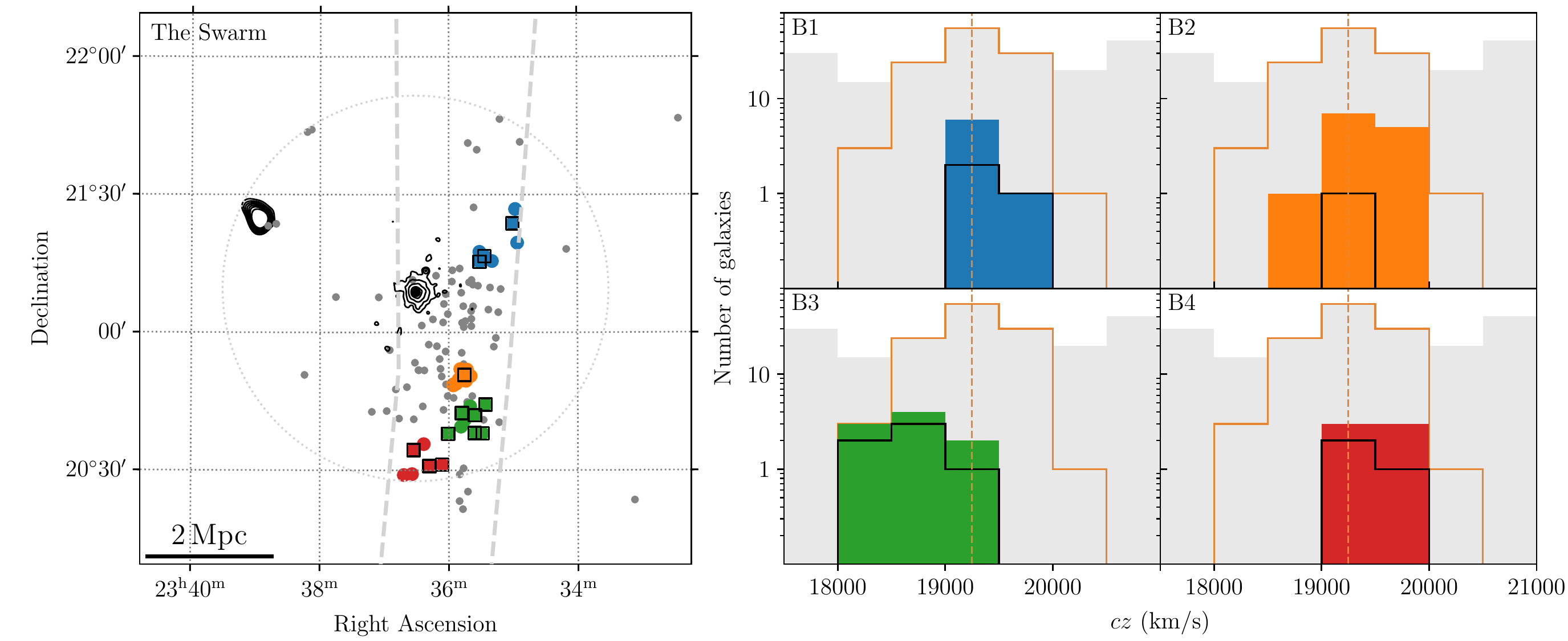}
        \includegraphics[width=.96\linewidth]{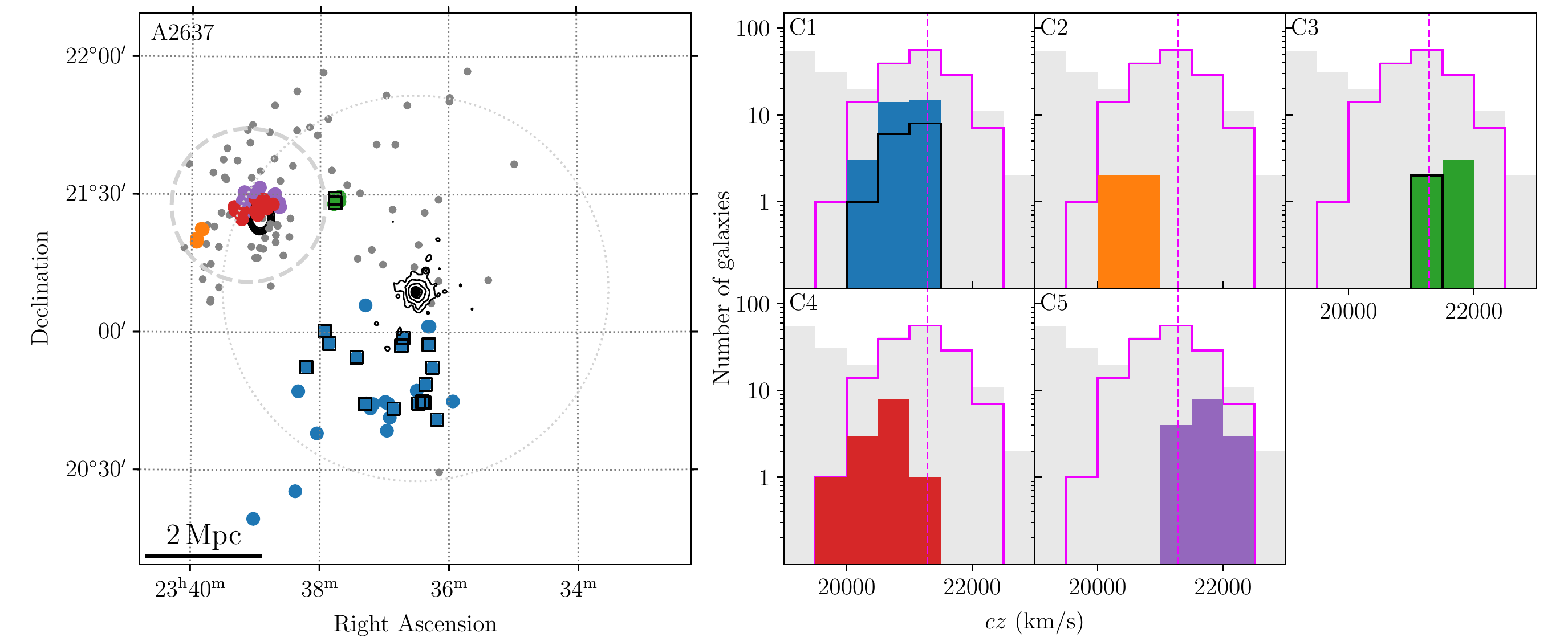}
        \vspace{-0.1cm}
        \caption{Sky distribution and velocity histograms of identified groups in A2626, the \hwone, and A2637. The large faint gray dotted lines show \rvirR of the clusters and overdensity. Contours show X-ray emission from \textit{ROSAT} at the center of A2626 and A2637. The groups identified using the DS test are indicated by the colored points. Black outlines indicate galaxies detected in \hi \citep{Healy2020b}. Gray points show galaxies associated with the overdensity but not part of any identified substructure. The velocity histograms of the identified groups are shown in the panels on the right. In the histogram panels, the background gray histogram shows the velocity distribution of the entire sample, and the open histograms outlined by green (A2626), orange (the \hwone), and pink (A2637) represent the galaxies identified as part of the clusters. The colored histograms correspond to the groups highlighted in the sky distribution plot. The black open histograms in each panel show the velocity distributions of the \hi-detected galaxies \citep{Healy2020b} in each group.}
        \label{fig:a2626_groups}
        \vspace{-1cm}
        \end{figure*}

        To identify groups of galaxies that are kinematically distinct from their parent cluster, we used the DS test, which has been successful in many other clusters \citep[e.g.,][]{Healy2020, Hess2015}. The DS test parameter 
        \begin{equation}\label{eqn:dstest}
        \delta_i^2 = \left( \frac{N_{\rm nn} + 1}{\sigma_{\rm cl}^2} \right) 
        \left[ ( \bar{v}_{\rm local}^i - \bar{v}_{\rm cl})^2 + 
        (\sigma_{\rm local}^2 - \sigma_{\rm cl}^2) \right],
        \end{equation} 
        where \scl is the cluster velocity dispersion, and $v_{\rm cl}$ is the mean velocity of the cluster (given in \tabref{tab:mkclusters}), $\sigma_{\rm local}$ and $v_{\rm local}$ are the velocity dispersion and mean velocity of a putative group with $N_{\rm nn}$ nearest neighbors. We used $N_{\rm nn} = 5,$ 10, 15, 20, and 25 and identified groups that consistently appear with multiple $N_{\rm nn}$ values.\\

        We ran the DS test on all three of the overdensities in the field. A simple $\pm 3\scl$ cut in velocity results in contamination from the neighboring overdensities, and we therefore assigned galaxies to the different overdensities based on their location in the distance--velocity phase space (\figref{fig:a2626_phasespace}). To assign galaxies, we widened the trumpet by doubling $M_{200}$ and selected all galaxies within the wider trumpet and having velocities within $3\,\scl$ of the cluster velocity.\\

        \figref{fig:a2626_dstest} shows six groups associated with A2626, four groups with the \hwone, and five groups associated with A2637. 
        More detailed sky distribution and velocity histogram plots of the identified groups are shown in \figref{fig:a2626_groups}. Information on whether \hi is detected or not in the group galaxies is useful here as groups with galaxies still containing detectable \hi are more likely to be recent additions to the cluster environment. 

        \subsection{Substructure in A2626}
        \label{sec:substructure_a2626}
        \begin{table}
        \caption{Groups identified in A2626 using the DS test}
        \label{tab:a2626_dsgroups}
        \begin{tabular}{cccc} \hline
        Group ID & $N_m$ & $\bar{v}$ & $\sigma_g$ \\
        &       & (\kms)    &  (\kms)    \\ \hline
        A1       &  17   &   16130   &    339     \\  
        A2       &  13   &   16868   &    151     \\  
        A3       &   9   &   16660   &    156     \\  
        A4       &  49   &   17051   &    402     \\  
        A5       &   7   &   16104   &    161     \\  
        A6       &   24  &   17216   &    685     \\  \hline
        \end{tabular}
        \end{table}

        The locations and velocity distributions of the six groups identified within A2626 are shown in \figref{fig:a2626_dstest}, and the details are given in \tabref{tab:a2626_dsgroups}. Four of the groups (A1, A2, A3, and A4) are located outside  \rvirR of the cluster and contain between 10 and 60\% \hi detections. This suggests these groups are on first infall into the cluster. The two smaller groups A2 and A3 have mean velocities similar to the cluster velocity (\numunit{16576}{\kms)} consistent with  being accreted from the wall in which A2626 is embedded and therefore moving in the plane of the sky. Groups A1 and A4 have mean velocities respectively lower and higher than the cluster velocity. This could mean that A1 is falling into the cluster from behind, while A4 is falling in from the front. While there are no known filaments connecting to A2626,  A2622 (another member of SCL~213) is nearby to the north of A2626. However A1 and A4 are coming in from the northeast and southwest respectively, suggesting, along with the high fraction of \hi detections, that these groups are not being accreted along a filament but rather from the field.

        The two groups located inside  \rvirR (A5 and A6) tell us something about the recent accretion onto the cluster. A5 still has 25\% of its galaxies detected in \hi, suggesting this is a group new to A2626. A5's velocity relative to the cluster suggests falling in from behind the cluster. Group A6 is near the cluster center (offset \numunit{250}{\text{kpc}} northwest) and has no \hi detections. Presumably it was accreted early.

        The overall picture that A2626 has undergone a merger \citep{Mohr1996, Mohr1997,Wong2008} is supported by the X-ray observations. \citet{Wong2008} found a significant change in the X-ray temperature at a radius of \numunit{260}{\text{kpc}} from the center of the cluster. They suggested an ongoing or previous merger as the likely explanation. The offset position of A6 from the center of the cluster is consistent with the echo of a group merging with the cluster center. The projected distance between A6 and the center of A2626 roughly is consistent with the radius of the change in X-ray temperature. A6 includes the emission-line galaxy IC~5337, which \citet{Wong2008} suggested is infalling from the west of the cluster.

        \subsection{Substructure in the \hwone}
        \label{sec:substructure_hw1}

        The \hwone, as discussed above, is probably not a cluster but rather a linear collection of galaxies and galaxy groups. The DS test identified four distinct groups in this overdensity. B3 might be on an infall trajectory towards A2626, its mean velocity being only $3.2\,\sigma_{A2626}$ higher than the cluster velocity of A2626 and located just beyond A2626's \rvirR. In this scenario, the higher relative mean velocity of B3 would suggest it is approaching A2626 from the front.  While most of B3 was not included in the DS test for A2626 due to the galaxy velocities being greater than $3\,\scl$ from A2626, we see hints of the eastern and lower-velocity side of this group in \figref{fig:a2626_dstest}b around $23^\mathrm{h}36^\mathrm{m}\,\,+20^\circ35'$).\\

        The three groups B1, B2, and B4 all have mean velocities that differ too much from A2626's for them to be associated. B1, B2, and B4 are also linearly aligned and have mean velocities that are within \numunit{150}{\kms}. However it is unclear whether this is a chance alignment or a result of the underlying large scale structure. B1, B3, and B4 include $>$40\% \hi detections, implying that these groups have been little-affected by stripping.

        \begin{table}
        \caption{Groups identified in the \hwone using the DS test}
        \label{tab:a2625_dsgroups}
        \begin{tabular}{cccc} \hline
        Group ID & $N_m$ & $\bar{v}$ & $\sigma_g$  \\
        &       & (\kms)    &  (\kms)     \\ \hline
        B1       &  7    &   19409   &    79       \\  
        B2       &  13   &   19433   &    282      \\  
        B3       &   9   &   18731   &    325      \\  
        B4       &   6   &   19560   &    248      \\  \hline
        \end{tabular}
        \end{table}

        \subsection{Substructure in A2637}
        \label{sec:substructure_a2637}

        Our substructure analysis of A2637 is not as complete as for A2626 or the \hwone because A2637 is on the edge of the MMT survey. This cluster is also outside of the MeerKAT FWHM, which means the \hi sensitivity for this cluster is not as good as for A2626 and the \hwone. Despite this, we identified four groups C2, C3, C4, and C5 associated with the cluster and one group, C1, at a similar redshift but \numunit{{>}1}{\mpc} from the cluster. C2 and C3 both appear to be infalling or recently accreted (particularly  C2). However, the paucity of \hi data leaves the gas-stripped fractions uncertain. \\

        There are two groups, C4 and C5, near the center of A2637. The two are similar in size but have mean velocities differing by \numunit{{>}1100}{\kms}. At the heart of C5 is 2MASXI~J2338533+212752,  identified as a brightest cluster galaxy (BCG)  \citep{Lauer2014}. Based on this, we hypothesize that C5 is at the center of the cluster, and C4 has been accreted and is merging with C5. Similar substructure is seen, for example, at the center of the Coma cluster, where there are separate X-ray emission peaks coinciding with the BCGs associated with two substructures \citep{Adami2005, Healy2020}. It is also widely accepted that the two groups at the center of the Coma cluster are merging \citep[e.g.,][]{Colless1996,Adami2005}. In the case of groups C4 and C5 in A2637, no such detailed X-ray analysis has yet been carried out.\\

        The diffuse group C1, which contains nearly 50\% \hi detections, does not appear to be associated with A2637. Its center is ${\sim} 3\,R_{200}$ from A2637, and there are no galaxies between C1 and A2637 which could hint at a connection. This raises the question of how C1 fits into the large scale structure of the system, but the limited field of view of the MMT survey and the lack of available redshifts outside it make it difficult to speculate on the possible origins of C1.

        \begin{table}
        \caption{Groups identified in A2637 using the DS test}
        \label{tab:a2637_dsgroups}
        \begin{tabular}{cccc} \hline
        Group ID & $N_m$ & $\bar{v}$ & $\sigma_g$  \\
        &       & (\kms)    &  (\kms)     \\ \hline
        C1       &  32   &   20905   &    242      \\  
        C2       &  4    &   20521   &    97       \\  
        C3       &   5   &   21527   &    180      \\  
        C4       &  13   &   20572   &    287      \\  
        C5       &   15  &   21755   &    329      \\  \hline
        \end{tabular}
        \end{table}

\section{Summary}
\label{sec:summary}

There are now over 2900 redshifts in a 2\degr$\times$2\degr\ field centered on the cluster A2626. The A2626 complex at is made up of A2626 itself ($z\approx 0.055$), the cluster A2637 ($z\approx0.071$), and an apparent wall or at least an extended structure designated here as the \hwone ($z\approx0.064$). The structure is linear rather than centrally concentrated, and its origins and connections to the large scale structure around the clusters remain unclear. On a larger scale, the A2626--The \hwone--A2637 complex appears not connected to the Perseus--Pegasus filament.

At much larger distances than A2637, the cluster A2625, formerly thought to be at roughly the same redshift as A2626 \citep{Struble1999}, is in fact the same cluster as RM J233602.7+203245.1 at $z\approx0.102$ (\numunit{cz = 30519}{\kms}). There are at least three and probably four other clusters in the background, the most distant at $z\approx 0.42$.

The greater numbers of confirmed members of A2626, A2637, and the \hwone have revealed substructures within these systems. There are six substructures within A2626, five within A2637, and probably four within the \hwone, although one of these last could instead be a group associated with A2626. The new redshifts have also decreased the uncertainties in the systems' velocity dispersions and sizes, which are now known to better than 10\%.

This work has shown the importance of extensive spectroscopy in identifying large-scale structure and linking components together. Even with the existing data, though, some questions remain: how is the \hwone connected to the structure around it? And how do A2626 and A2637 connect to the other members of the supercluster SCL~213? Answering these questions will require spectra over a still wider area on the sky than observed here, and we look to the future of wide-area spectroscopic surveys such as the WEAVE Wide-field Cluster Survey (Jin et al.\ 2021, in prep.)\ to be able to answer some of the outstanding questions about the environment in which A2626 is embedded.  Wider-field \hi observations will also be valuable.

\acknowledgements

Thank you to the anonymous referee for their comments that have improved this paper.

JH acknowledges research funding from the South African Radio
Astronomy Observatory, which is a facility of the National Research
Foundation, an agency of the Department of Science and Innovation. We
thank K Hess for useful discussions about groups and filaments. We
thank the UCT Writing Circle for suggestions that improved the
readability of this work.

MV acknowledges support by the Netherlands Foundation for Scientific
Research (NWO) through VICI grant 016.130.338.

This paper uses data products produced by the OIR Telescope Data
Center, supported by the Smithsonian Astrophysical Observatory. The
authors thank Nelson Caldwell for help with the Hectospec
configuration design and queue scheduling processes, Sean Moran for
maintaining the Hectospec data pipeline, and Jessica Mink for help
with {\sc xcsao}.

This research has made use of the SIMBAD database, operated at CDS,
Strasbourg, France.

Funding for the Sloan Digital Sky Survey IV has been provided by the
Alfred P. Sloan Foundation, the U.S. Department of Energy Office of
Science, and the Participating Institutions. SDSS-IV acknowledges
support and resources from the Center for High-Performance Computing
at the University of Utah. The SDSS web site is
\url{www.sdss.org}. SDSS-IV is managed by the Astrophysical Research
Consortium for the Participating Institutions of the SDSS
Collaboration including the Brazilian Participation Group, the
Carnegie Institution for Science, Carnegie Mellon University, the
Chilean Participation Group, the French Participation Group,
Harvard-Smithsonian Center for Astrophysics, Instituto de
Astrof\'isica de Canarias, The Johns Hopkins University, Kavli
Institute for the Physics and Mathematics of the Universe
(IPMU)/University of Tokyo, the Korean Participation Group, Lawrence
Berkeley National Laboratory, Leibniz Institut f\"ur Astrophysik
Potsdam (AIP), Max-Planck-Institut f\"ur Astronomie (MPIA
Heidelberg), Max-Planck-Institut f\"ur Astrophysik (MPA Garching),
Max-Planck-Institut f\"ur Extraterrestrische Physik (MPE), National
Astronomical Observatories of China, New Mexico State University, New
York University, University of Notre Dame, Observat\'ario
Nacional/MCTI, The Ohio State University, Pennsylvania State
University, Shanghai Astronomical Observatory, United Kingdom
Participation Group, Universidad Nacional Aut\'onoma de M\'exico,
University of Arizona, University of Colorado Boulder, University of
Oxford, University of Portsmouth, University of Utah, University of
Virginia, University of Washington, University of Wisconsin,
Vanderbilt University, and Yale University.

%



\software{astropy \citep{Astropy2013, Astropy2018},  
            Matplotlib \citep{Hunter2007},
            iraf \citep{Tody1986,Tody1993},
            xcsao \citep{Mink2007}
          }





\bibliographystyle{aasjournal}
\bibliography{references}{}



\end{document}